\def\aa{{\it Astron. Astrophys.\/} }

\def\aj{{\it Astron. J.\/} }
\def\annrev{{\it Annu. Rev. Astr. Astrophys.,} }
\def\apj{{\it Ap. J.\/} }
\def\apjl{{\it Ap. J. Lett.\/} }
\def\apjs{{\it Ap. J. Suppl.\/} }     
      
\def\mnras{{\it MNRAS} }
\def\nat{{\it Nature\/} }

\def\PASP{{\it Publ. Astr. Soc. Pacif.,} }

\def\ptr{{\it Phil. Trans. R. Soc. Lond. A\/} }

\input psfig.sty

\documentstyle[epsfig]{arpp}

\begin{document}
\title{MODIFIED NEWTONIAN DYNAMICS AS AN ALTERNATIVE TO DARK MATTER}
\author{Robert H.\ Sanders$^1$ 
\& Stacy S.\ McGaugh$^2$
\affiliation{$^1$Kapteyn Astronomical Institute,
University of Groningen, Groningen, The Netherlands \\
$^2$Department of Astronomy, University of Maryland, College
Park, MD, USA}}
\markboth{}{Sanders \& McGaugh --- MODIFIED NEWTONIAN DYNAMICS}

\begin{abstract}

Modified Newtonian dynamics (MOND) is an empirically motivated modification
of Newtonian gravity or inertia suggested by Milgrom as an alternative
to cosmic dark matter.  The basic idea is that at accelerations below
$a_o\approx 10^{-8}$ cm/s$^2$ $\approx cH_o/6$ 
the effective gravitational attraction approaches $\sqrt{g_na_o}$ where
$g_n$ is the usual Newtonian acceleration.
This simple algorithm yields flat rotation 
curves for spiral galaxies and a mass-rotation velocity relation of the form 
$M\propto V^4$ that forms the basis for the observed luminosity-rotation 
velocity relation-- the Tully-Fisher law.  We review the phenomenological
success of MOND on scales ranging from dwarf spheroidal galaxies to 
superclusters, and demonstrate that the evidence for dark matter can
be equally well interpreted as evidence for MOND.  We discuss the possible
physical basis for an acceleration-based modification of Newtonian dynamics
as well as the extension of MOND to cosmology and structure formation.  
\end{abstract}

\maketitle

\section{INTRODUCTION}

The appearance of discrepancies between the Newtonian dynamical mass
and the directly observable mass in large astronomical systems has two possible
explanations:  either these systems contain large quantities of unseen
matter, or gravity (or the response of particles to gravity) 
on these scales is not described by Newtonian theory.  Most attention
has focused on the first of these explanations.  An intricate
paradigm has been developed in which non-baryonic dark matter plays a 
central role-- not only in accounting for the traditional dynamical mass 
of bound gravitational systems (Faber \& Gallagher 1979, Blumenthal et al
1984) but also in promoting the formation
of structure via gravitational collapse beginning in the highly homogeneous
ionized universe (Peebles 1982, Vittorio \& Silk 1984).  
The paradigm of cold dark matter (CDM) is widely
purported to be successful in this cosmological context, particularly 
in predicting the scale-dependence of density fluctuations.  Moreover, with 
the development
of cosmic N-body simulations of high precision and resolution, this hypothesis
has gained predictive power on the scale of galaxies-- a power that 
considerably restricts the freedom to arrange dark matter as one would
wish in order to explain the form and magnitude of the discrepancy in any
particular system (Navarro et al. 1996).

It is in this second aspect, the distribution of dark matter in galactic
systems, that the CDM paradigm encounters observational difficulties
(McGaugh \& de Blok 1998a, Sellwood \& Kosowsky 2001). It is not the purpose
of this review to discuss the possible problems with the CDM hypothesis. We
only comment that, as of the date of this review, candidate dark matter
particles have not yet been detected by any means independent of their
putative global gravitational effect.  So long as this is the only evidence
for dark matter, then its presumed existence is not independent of the assumed
law of gravity or inertia on astronomical scales.  If a physical law, when
extended to a  regime in which it has never before been
tested, implies the existence of a medium (e.g. an ether) that cannot be
detected by any other means, then it would not seem 
unreasonable to question that law.
 
Of course, if one chooses to modify Newtonian dynamics or gravity in an
ad hoc fashion, then the set of alternative
possibilities is large.  It is a simple matter to claim that 
Newton's law of gravity fails on galactic scales and then to cook up
a recipe that explains a particular aspect of the observations-- such
as flat rotation curves of spiral galaxies.  To be credible, an
empirically based alternative to dark matter should at least provide a more
efficient description of the phenomenology.  Any viable alternative
should account for various aspects of the observations of astronomical
systems (such as global scaling relations) with as few additional parameters
as possible.  A second, but less immediate, requirement is that the
suggested alternative should have some basis in sensible physics-- it
should make contact with familiar physical principles or at least a 
reasonable extrapolation of those principles.

To date, the only suggestion that goes some way toward meeting these
requirements (particularly the former) is Milgrom's modified Newtonian 
dynamics
(Milgrom 1983 a,b,c): MOND.  The empirical successes of this hypothesis on
scales ranging from dwarf spheroidal galaxies to super-clusters, and
its possible physical basis and extension to a cosmological context, is the
subject of this review.  It may be argued that this is a speculative
topic for review in this series.  In our opinion, the subject of
dark matter (Trimble 1987) is, in the absence of its direct detection, no 
less speculative-- particularly considering that the
standard model of particle physics does not predict the existence of 
candidate dark matter particles with the necessary properties.  
Reasonable extensions of the standard model (e.g. supersymmetry) can, with an
appropriate adjustment of parameters, accommodate such particles, but this
also requires an extrapolation of known physics (e.g., Griest, et al. 1990).
Here we demonstrate that the evidence for dark matter 
can be equally well interpreted as evidence for modified dynamics.

There have been other attempts to modify gravity
in order to account for astronomical mass discrepancies without invoking dark
matter. We will mention some of these efforts, but it is fair to say that
none of these alternatives has enjoyed the phenomenological success of
MOND or been as extensively discussed in the literature.
A considerable lore on  MOND has emerged in the past two decades-- with
contributions not only by advocates reporting phenomenological successes but
also by critics pointing out possible problems or inconsistencies.  There
have also been several contributions attempting to formulate MOND either
as a covariant theory in the spirit of General Relativity, or as a modified
particle action (modified inertia).  Whereas none of these attempts has, so
far, led to anything like a satisfactory or complete theory, they
provide some insight into the required properties of generalized theories of
gravity and inertia.  

In the absence of a complete theory MOND cannot be unambiguously extended
to problems of cosmology and structure formation.  However, by making
certain reasonable assumptions, one may speculate on the form of a
MOND cosmology.  We discuss the efforts that have been made in this
direction.  The general expectation is that, because MOND results in
effectively stronger gravity for low peculiar accelerations, the rapid
growth of structure is possible even in a low-density purely baryonic 
Universe.

\section{BASICS OF Modified Newtonian Dynamics}
\subsection {An acceleration scale}
The phenomenological basis of MOND consists of two
observational facts about spiral galaxies: 1.  The rotation curves
of spiral galaxies are asymptotically flat (Shostak 1973,
Roberts \& Whitehurst 1975, Bosma 1978, Rubin et al 1980), and 2. 
There is a well-defined relationship between the rotation velocity in spiral
galaxies and the luminosity-- the Tully-Fisher (TF) law (Tully \& Fisher 1977,
Aaronson et al. 1982).  This latter implies a mass-velocity   
relationship of the form $M\propto V^\alpha$ where $\alpha$ is in the
neighborhood of 4.  

If one wishes to modify gravity in an {\it ad hoc} way to explain 
flat rotation curves, an
obvious first choice would be to propose that gravitational attraction becomes
more like 1/r beyond some length scale which is comparable to the scale of
galaxies.  So the modified law of attraction about a point mass M would read
$$F = {GM\over{r^2}}{f(r/r_o)}\eqno(1)$$ where $r_o$ is a new constant of 
length on the order of a few kpc, and $f(x)$ is a function with the
asymptotic behavior: $f(x)=1$ where $x\ll 1$ and $f(x)=x$ where $x\gg 1$. 
Finzi (1963), Tohline (1983), Sanders (1984)
and Kuhn \& Kruglyak (1987) have proposed variants of this idea.  In Sanders'
(1984) version, the Newtonian potential is modified by including a repulsive
Yukawa term ($e^{-r/r_o}/r$) which can yield a flat rotation velocity over
some range in $r$. This idea keeps re-emerging with various modern
justifications; e.g., Eckhardt (1993), Hadjimichef \& Kokubun (1997), Drummond
(2001) and  Dvali et al. (2001).  

All of these modifications attached to a length scale have one thing in
common: equating the centripetal to the
gravitational acceleration in the limit $r > r_o$  leads to a  
mass-asymptotic rotation velocity relation of the form $v^2 = GM/r_o$.  
Milgrom (1983a) realized that this was
incompatible with the observed TF law, $L\propto v^4$.  
Moreover, any modification attached to a length
scale would imply that larger galaxies should exhibit a larger discrepancy
(Sanders 1986).  This is contrary to the observations.
There are very small, usually
low surface brightness (LSB) galaxies with large discrepancies, and very 
large high surface brightness (HSB) spiral
galaxies with very small discrepancies (McGaugh \& de Blok 1998a).  

Figure 1 illustrates this.
At the left is a log-log plot of the dynamical $M/L_{K'}$
vs.\ the radius at the last measured
point of the rotation curve for a uniform sample of spiral 
galaxies in the Ursa Major cluster (Tully et al. 1996, 
Verheijen and Sancisi, 2001).  The dynamical
M/L is calculated simply using the Newtonian formula for the mass $v^2r/G$ 
(assuming a spherical mass distribution) where r is the 
radial extent of the rotation curve.
Population synthesis  studies suggest that $M/L_{K'}$ should be about one,
so anything much above one indicates a global discrepancy-- 
a ``dark matter problem''.  
It is evident that there is not much of a correlation of M/L with size.  
On the other
hand, the Newtonian M/L plotted against centripetal acceleration ($v^2/r$)
at the last measured point (Figure 1, right) looks rather different.  
There does appear to be a correlation in the sense that $M/L\propto 1/a$ for
$a<10^{-8}$ cm/s$^2$.  The presence of an acceleration scale in the
observations of the discrepancy in spiral galaxies has been pointed out
before (Sanders 1990, McGaugh 1998), and, as the data have improved, it 
has become more evident.  Any modification of gravity  attached to a length
scale cannot explain such observations.

Milgrom's insightful deduction was that the only viable sort of modification
is one in which a deviation from Newton's law appears at low acceleration.
(Data such as that shown in Fig.\ 1 did not exist
at the time of Milgrom's initial papers; an acceleration scale was indicated
by the slope of the TF relation.)  
MOND as initially formulated could be viewed as a modification of 
inertia or of gravity (this dichotomy remains).  In the first case, the 
acceleration of a particle with mass $m$ under the influence of an
external force, would be given by $$m{\bf a}\mu(a/a_o) = {\bf F}\eqno(2)$$
where $a_o$ is a new physical parameter with
units of acceleration and $\mu(x)$ is a function that is unspecified
but must have the asymptotic form $\mu(x) = x$ when $x\ll 1$ and 
$\mu(x) = 1$ when $x\gg 1$. 
Viewed as a modification of gravity, the true
gravitational acceleration ${\bf g}$ is related to the Newtonian gravitational
acceleration ${\bf g_n}$ as $$ {\bf g}\mu(|g|/a_o) = {\bf g_n}.\eqno(3)$$ 

Although there are clear differences in principle and practice between these
two formulations, the consequence for test particle motion in a gravitational
field in the low
acceleration regime is the same:  the effective gravitational force becomes:
$g = \sqrt{g_na_o}$.  For a point mass M, if we set $g$ equal to the 
centripetal acceleration $v^2/r$, this gives $$v^4 = GMa_o\eqno(4)$$ in the 
low acceleration regime.   Thus, all rotation curves of isolated masses are
asymptotically flat and there is a mass-luminosity relation of the form
$M\propto v^4$.  These are aspects that are built into MOND so they
cannot rightly be called predictions.  However,
in the context of MOND, the  aspect of an asymptotically flat rotation curve 
is absolute.  MOND leaves rather little room for maneuvering;  the idea is 
in principle
falsifiable, or at least it is far more fragile than the 
dark matter hypothesis.  Unambiguous examples of rotation curves 
of isolated galaxies which 
decline in a Keplerian fashion at a large distance
from the visible object would falsify the idea. 

In addition, the mass-rotation velocity relation and implied Tully-Fisher
relation is absolute.  The
TF relation should be the same for different classes of galaxies, 
independent of length scale or surface brightness, and the
logarithmic slope (at least of the mass-velocity relation) must be 4.0.
Moreover, the relation is essentially one between the total
baryonic mass of a galaxy and the asymptotic flat rotational velocity--
not the peak rotation velocity but the velocity at large distance.
This is the most immediate and most obvious prediction (see McGaugh \&
de Blok 1998b and McGaugh et al 2000 for a discussion of these points).  

Converting the M-V relation (eq.\ 4)
to the observed luminosity-velocity relation we find
$$ \log(L) = 4\log(v) - \log(Ga_o<M/L>).\eqno(5)$$
Figure 2 shows the near-infrared TF relation for Verheijen's UMa sample
\cite{sv98} where the rotation velocity, $v$, is that of
the flat part of the rotation curve.  The scatter about the 
least-square fit line of slope $3.9 \pm 0.2$ is consistent with observational
uncertainties (i.e., no intrinsic scatter).
Given the mean M/L in a particular band ($\approx 1$ in the 
K' band), this observed TF relation (eq.\ 5) tells us that
$a_o$ must be on the order of $10^{-8}$ cm/s$^2$.  
Milgrom immediately noticed
that $a_o\approx cH_o$ to within a factor of 5 or 6.
This cosmic coincidence is provocative and suggests that MOND
perhaps reflects the effect of cosmology on local particle dynamics.

\subsection{General predictions}

There are several other direct observational consequences of modified 
dynamics--
all of which Milgrom explored in his original papers-- that
are genuine predictions in the sense that they are not part
of the propositional basis of MOND.

1.  There exists a critical value of the surface density
$$\Sigma_m \approx a_o/G.\eqno(6)$$  If a system, such as a spiral galaxy has
a surface density of matter greater than $\Sigma_m$, then
the internal accelerations are greater than $a_o$, so the system is
in the Newtonian regime.  In systems with $\Sigma \geq \Sigma_m$ (HSB
galaxies) there should be a small discrepancy between
the visible and classical Newtonian dynamical mass within the optical disk. 
In the parlance of rotation curve observers, a HSB galaxy should be
well-represented by the ``maximum disk'' solution (van Albada \& Sancisi 
1986).  But in LSB galaxies ($\Sigma\ll \Sigma_m$) there is a low internal 
acceleration, so the discrepancy between the visible
and dynamical mass would be large.  Milgrom predicted, before the actual
discovery of LSB galaxies, that there should be a serious
discrepancy between the observable and dynamical mass within the
luminous disk of such systems-- 
should they exist.  They do exist, and
this prediction has been verified-- as is evident from the work of
McGaugh \& de Blok (1998a,b) and de Blok \& McGaugh (1998).

2.  It is well-known since the work of Ostriker \& Peebles (1973) that 
rotationally supported Newtonian systems tend to be unstable to
global non-axisymmetric modes that lead to bar formation and rapid
heating of the system.  In the context of MOND, these systems would be
those with $\Sigma > \Sigma_m$, so this would suggest that $\Sigma_m$
should appear as an upper limit on the surface density of rotationally 
supported systems.  This critical surface density is
0.2 g/cm$^2$ or 860 M$_\odot$/pc$^2$.  A more appropriate value of the mean
surface density within an effective radius would be $\Sigma_m/2\pi$
or 140 M$_\odot/pc^2$, and, taking
$M/L_b \approx 2$, this would correspond to a surface brightness of
about 22 mag/arc sec$^2$.  There is 
such an observed upper limit on the mean surface brightness of spiral 
galaxies and this is known as
Freeman's law (Freeman 1970, Allen \& Shu 1979).  The point is that 
the existence of such a maximum surface density (McGaugh et al. 1995b, McGaugh
1996) is quite natural in the context of MOND, but must be put in by hand
in dark matter theories (e.g. Dalcanton et al. 1997).

3.  Spiral galaxies with a mean surface density near this limit -- 
HSB galaxies-- would be, within
the optical disk, in the Newtonian regime.  So one would expect that
the rotation curve would decline in a near Keplerian fashion to the 
asymptotic constant value.  In LSB galaxies, with mean surface density
below $\Sigma_m$, the prediction is that rotation curves continuously rise to
the final asymptotic flat value.  Thus, there should be a general difference
in rotation curve shapes between LSB and HSB galaxies.
Fig.\ 3  shows the rotation curves of two galaxies, a LSB and HSB, 
where we
see exactly this trend.  This general effect in observed rotation curves
was first noted by Casertano \& van Gorkom (1991).

4.  With Newtonian dynamics, pressure-supported systems that 
are nearly isothermal have infinite mass.  However, in the context of MOND
it is straightforward to demonstrate that such isothermal systems
are of finite mass with the density at large radii falling approximately as
$r^{-4}$ (Milgrom 1984).  
The equation of hydrostatic equilibrium for an isotropic, isothermal
system reads $${\sigma_r}^2 {{d\rho}\over{dr}} = -\rho g\eqno(7)$$
where, in the limit of low accelerations $g=\sqrt{GM_r a_o}/r$.
$\sigma_r$ is the radial velocity dispersion, $\rho$ is
the mass density, and $M_r$ is the mass enclosed within r.
It then follows immediately that, in the outer regions, where
$M_r = M$ = constant, 
$$\sigma_r^4 = {GMa_o\Bigl({{d\,ln(\rho)}\over{d\,ln(r)}}\Bigl)^{-2}}.
\eqno(8)$$
Thus there exists a mass-velocity dispersion
relation of the form $$(M/10^{11}M_\odot) \approx (\sigma_r/100\,\,
kms^{-1})^4\eqno(9)$$ which is similar to the observed
Faber-Jackson relation (luminosity-velocity dispersion relation) 
for elliptical galaxies \cite{fj}.   
This means that a MOND near- isothermal sphere 
with a velocity dispersion of 100 km/s to
300 km/s will always have a galactic mass. 
Moreover, the same $M-\sigma$ relation 
(eq.\ 8) should apply to all pressure supported, near-isothermal systems, 
from globular clusters to clusters of galaxies, albeit with considerable
scatter owing to deviations from a strictly isotropic, isothermal
velocity field (Sanders 2000).  
 
The effective radius of a near-isothermal system is roughly
$r_e\approx \sqrt{GM/a_o}$ (at larger radii the system is in the
MOND regime and is effectively truncated).  This means that $a_o$ 
appears as a characteristic acceleration in near-isothermal systems
and that $\Sigma_m$ appears
as a characteristic surface density-- at least as an upper limit
(Milgrom 1984).
Fish (1964), on the basis of then existing photometry, pointed out 
that elliptical galaxies exhibit 
a constant surface brightness within an effective radius.  
Although Kormendy (1982) demonstrated that more luminous ellipticals have 
a systematically lower surface brightness, when the ellipticals are considered 
along with the bulges of spiral galaxies and globular 
clusters (Corollo et al 1997), there does appear to be a characteristic 
surface brightness which is on the order of that implied by 
$\Sigma_m$, i.e., the Fish law is recovered for the larger set of
pressure supported objects.

5.  The ``external field effect" is not a prediction but
a phenomenological requirement on MOND that has strong implications
for non-isolated systems.  In his original papers, Milgrom noted that open star
clusters in the Galaxy do not show evidence for mass discrepancies
even though the internal accelerations are typically below $a_o$.  He
therefore postulated that the external acceleration field of the Galaxy must
have an effect upon the internal dynamics of a star cluster-- that, in
general, the dynamics of a any sub-system is affected by the external field in
which that system is found.  This implies that the theory upon which MOND is
based does not respect the equivalence principle in its strong form (this in
no way implies that MOND violates the universality of free fall-- the weak
version of the equivalence principle-- which is the more cherished and
experimentally constrained version).  Milgrom suggested that this effect arises
owing to the non-linearity of MOND and can be approximated by including the
external acceleration field, ${\bf g_e}$ in the MOND equation, i.e.,
$$\mu(|{\bf g_e} + {\bf g_i}|/a_o){\bf g_i} = {\bf g_{n_i}}\eqno(10)$$  
where ${\bf g_i}$ is the internal
gravitational field of the sub-system and ${\bf g_{n_i}}$ is the Newtonian 
field of the sub-system alone.  This means that a sub-system with internal
accelerations below $a_o$ will exhibit Newtonian dynamics if the external
acceleration exceeds $a_o$ ($g_i<a_o<g_e$).  If the external and 
internal accelerations are
below $a_o$ and $g_i<g_e<a_o$, 
then the dynamics of the subsystem will be Newtonian but with a
larger effective constant of gravity given by $G/\mu(g_e/a_o)$.  
If $g_e<g_i<a_o$ the dynamics is MONDian but with a maximum discrepancy 
given by $[\mu(g_e/a_o)]^{-1}$.  In addition
the dynamics is anisotropic with  dilation along the direction of the external
field.

The external field effect would have numerous consequences:  it would  
influence the internal dynamics
of globular clusters and the satellite dwarf companions of the Milky Way
independently of tides.   It may provide
an important non-tidal mechanism for the maintenance of warps in galaxies 
owing to the presence of companions.  The peculiar
accelerations resulting from large scale structure would be expected to limit
the mass discrepancy in any particular galactic system (no object is isolated),
and the deceleration (or acceleration) of the Hubble flow may influence the
development of large scale structure.

\subsection{Dark halos with an acceleration scale}

Can the phenomenology of MOND be reproduced by dark halos-- specifically
the kind of dark halos that emerge from cosmological N-body simulations 
with an initial fluctuation spectrum given by cold dark matter (CDM)?  This is
an important question because the phenomenological success of MOND may 
be telling us something about the universal distribution of dark matter in
galaxies and its relation
to the visible component rather than anything about the law of gravity or
inertia at low accelerations.  This question was first considered by Begeman
et al.\ (1991) who attempted to devise disk-halo coupling rules that could
yield a one-parameter fit to rotation curves (M/L of the visible disk) similar
to modified Newtonian dynamics (MOND).  Without any physical justification the
core radius and asymptotic circular velocity of an isothermal halo were
adjusted to the scale length and maximum rotation velocity of the disk to
yield a  characteristic acceleration.  With such coupling rules, the fits
to galaxy rotation curves were of lower quality than the MOND fits 
(particularly for the dwarf systems), and there were
numerous ambiguities (e.g., in gas dominated galaxies, what is the
proper disk length scale?).  Similar {\it ad hoc} coupling 
rules between visible and dark components have also been considered by 
Giraud (2000).

The idea that the halo might exhibit a characteristic acceleration was 
carried further by Sanders \& Begeman (1994) when the first cosmic N-body
calculations with high resolution (Dubinski \& Carlberg 1991) 
indicated that CDM halos were not
at all similar to an isothermal sphere with a constant density core. 
The objects emerging from the simulations
exhibited a density law with a $r^{-1}$ cusp that steepened in the outer 
regions to $r^{-4}$.  Dubinski \& Carlberg pointed out that this run of 
density was well described by the model of Hernquist (1990):
$$\rho(r) = {{\Sigma_o}\over r}(1 + r/r_o)^{-3}\eqno(11)$$ 
This has been subsequently confirmed by the extensive calculations of
Navarro et al.\ (1997), who corrected the outer power law to -3
(this is the famous NFW halo).  The reality of the cusp, if not the exact
power law, seems well-established (Moore et al 1998),
and is due to the fact that there are no phase-space constraints upon the
density of a collapsed object composed of CDM.

Sanders \& Begeman (1994) pointed out that if $\Sigma_o$ were 
fixed with only the
characteristic length scale varying from halo to halo, this implied
that a fixed acceleration scale could be associated with any halo,
$a_h = 2\pi G\Sigma_o$.  They demonstrated that a halo density law of this
form provided a reasonable fit to rotation curves of several high surface
brightness (HSB)
galaxies (comparable to that of MOND) where the length scale of the halo was 
proportional to the mass of the visible disk.  This proportionality would
follow if the baryonic mass were a fixed fraction (about 0.03) 
of the dark mass as is usually assumed, so this appeared to be
a natural way to explain MOND phenomenology in the context of CDM.

There are two problems with this idea:  First, a fixed
$\Sigma_o$ implies that no galaxy could exhibit an acceleration in
the inner regions less than $a_o$;  this is not true for a number of
low surface brightness (LSB) galaxies (McGaugh \& de Blok 1998b).  The problem
was already evident in the fits to the LSB galaxies in the sample of Sanders \&
Begeman in which the one-parameter fitting scheme broke down.  Secondly, the
halos that emerge from the cosmic N-body simulations do not have fixed
$\Sigma_o$ as is evident from the mass-rotation velocity law of $m\propto v^3$
(NFW).  No  characteristic surface density or acceleration is evident in these
objects.

Semi-analytic models for the formation of disk galaxies in the
context of CDM (van den Bosch \& Dalcanton 2000) can be tuned to give
rise to a characteristic acceleration.
In such models, one starts with a specified dark halo density law (NFW), 
allows some fraction of the halo mass, presumably baryonic, to collapse 
by a factor determined
by the dimensionless spin parameter of the halo, applies a stability
criterion to allow some further fraction of this dissipational component 
to be converted to stars, and removes gas from the system by 
an appropriate number of supernovae (feedback).  In this procedure there are
dimensionless parameters which quantify the feedback mechanism.  Because 
these parameters can be
adjusted to produce a Tully-Fisher (TF) law of the form $L\propto V^4$, 
then it is not surprising that there is
a fixed acceleration connected with these models ($a \approx V^4/GL<M/L>$).
The exercise is essentially that of modeling 
complicated astrophysical processes by a set of free parameters;  
these parameters are then tuned in order to achieve a desired result.
The fact that this is possible is interesting, but not at all compelling.

The possibility of a characteristic acceleration arising from CDM 
has been revisited
by Kaplinghat \& Turner (2002), who offered an explanation for why dark matter 
appears to become dominant beyond an acceleration numerically comparable
to $cH_o$.  They argued that halos formed from CDM possess
a one-parameter density profile that leads to a characteristic acceleration 
profile that is only weakly dependent upon the mass (or comoving scale) of 
the halo.  Then with a fixed collapse factor for the baryonic material, 
the transition from dominance of dark over baryonic occurs at a universal 
acceleration which, by numerical coincidence, is on the order of $cH_o$.  
Milgrom (2002) responded by pointing out that $a_o$ plays several roles
in the context of MOND:  it not only is a transition acceleration below which
the mass discrepancy appears, but it also defines the asymptotic rotation
velocity of spiral galaxies (via eq.\ 4) and thereby determines the
zero point of the TF relation (eq.\ 5); $a_o$ 
determines the upper limit on
the surface density of spirals (i.e., $\Sigma_m$); $a_o$ appears as an
effective upper limit upon the gravitational acceleration of a halo component
in sensible disk-halo fits to observed rotation curves 
(Brada \& Milgrom 1999a);
$a_o$ determines the magnitude of the discrepancy within LSB galaxies
(where the ratio of missing to visible mass 
is $a_o/g_i)$; $a_o$ sets the scale of
the Faber-Jackson relation via eq.\ 8; $a_o$ appears as an effective internal
acceleration for pressure-supported, quasi-isothermal systems,
and determines the dynamics of galaxy systems-- global and detailed-- 
ranging from small groups
to super-clusters.  These roles are independent in the context of dark matter,
and would each require a separate explanation.  The explanation of Kaplinghat
\& Turner applies only to the first of these and by construction prohibits
the existence of objects which are dark matter-dominated within their
optical radius (such as LSB galaxies).

The basic problem in trying to explain a fixed acceleration scale in galaxies
in terms of galaxy formation rather than 
underlying dynamics is that the process is stochastic:  each galaxy has its
own history of formation-evolution-interaction.  One would expect these
effects to erase any intrinsic acceleration scale, not enhance it.
Dark matter may address the
general trends but it cannot account for the individual idiosyncrasies of
each rotation curve.  In the next section we present the evidence that MOND 
can do this with a single value of $a_o$.

\section{ROTATION CURVES OF SPIRAL GALAXIES}

\subsection{Method and results of rotation curve fitting}
The measured rotation curves of spiral galaxies constitute the ideal
body of data to confront ideas such as MOND (Begeman, Broeils \& Sanders
1991, Sanders 1996, McGaugh \& de Blok 1998b, Sanders \& Verheijen 1998). 
That is because, in the absence of dark matter, the rotation curve is
in principle predictable from the observed distribution of stars and gas.
Moreover,
the rotation curve as measured in the 21 cm line of neutral hydrogen often 
extends well beyond the optical image of the galaxy where the centripetal 
acceleration is small and the discrepancy is large.  In the particularly 
critical case of the LSB galaxies,
21 cm line observations can be supplemented by H$_\alpha$ observations
(McGaugh et al. 2001, de Blok \& Bosma 2002)
and compared in detail with the rotation curve predicted from the
distribution of detectable matter.
The procedure that has usually been followed is outlined below:

1.  One assumes that light traces stellar mass, i.e., mass-to-light ratio 
(M/L) is constant in a  given galaxy.  There
are color gradients in spiral galaxies so this cannot be generally
true-- or at least one must decide which color band is the best tracer
of the mass distribution.  The general opinion is that
the near-infrared emission of spiral galaxies is the optimal tracer
of the underlying stellar mass distribution because the old population of
low mass stars
contribute to this emission and the near-infrared is less affected by dust
obscuration.  Thus, where available, near infrared surface photometry is 
preferable.

2.  In determining the distribution of detectable matter one must 
include the observed neutral hydrogen scaled up with an appropriate 
correction factor (typically 1.3 to 1.4) to account for the
contribution of primordial helium.  The gas can make a dominant
contribution to the total mass surface density in some (generally low
luminosity) galaxies.  

3.  Given the observed distribution of mass, the Newtonian gravitational
force, $g_n$, is calculated via the classical Poisson
equation.  Here it is usually assumed that the stellar and gaseous disks
are razor thin.  It may also be necessary to include a spheroidal bulge 
if the light distribution indicates the presence of such a component.

4.  Given the radial distribution of the Newtonian force, the effective
gravitational force, $g$, is calculated from
the MOND formula (eq.\ 3) with $a_o$ fixed.    
Then the mass of the stellar disk is
adjusted until the best fit to the observed rotation curve is achieved.  This
gives M/L of the disk as the single free parameter of the fit (unless a bulge
is present).

In this procedure one assumes that the motion of 
the gas is co-planer rotation about the center of the given galaxy.  
This is certainly not always the case because there are well-known 
distortions to the velocity field in spiral galaxies caused by bars and 
warping of the gas layer.  In a fully 2-dimensional velocity field these 
distortions can often be modeled (Bosma 1978, Begeman 1989), 
but the optimal rotation curves are those 
in which there is no evidence for the presence of significant deviations 
from co-planar circular motion.  
Not all observed rotation curves are perfect tracers of the radial
distribution of force.  A perfect theory will not fit all rotation curves
because of these occasional problems (the same is true of a specified
dark matter halo).
The point is that with MOND, usually, there is one adjustable parameter per 
galaxy and that is the mass or M/L of the stellar disk.

The empirical value of $a_o$ has been derived from a highly selected
sample of large galaxies with well-determined rotation curves (Begeman et al.
1991).  Assuming a distance scale, $H_o = 75$ km/(s Mpc)
in this case, rotation curve fits to all galaxies in the sample were
made allowing $a_o$ to be a free parameter.  The mean value for nine
of the galaxies in the sample, excluding NGC 2841 with a distance 
ambiguity (see below), is $1.2\pm 0.27 \times 10^{-8}$ cm/s$^2$.
Having fixed $a_o$ in this way, one is no longer free to take it
as a fitting parameter.  

There is, however, a relation between
the derived value of $a_o$ and the assumed distance scale
because the implied centripetal acceleration in a galaxy scales as 
the inverse of the assumed distance.  With respect to galaxy rotation
curves, this dependence is not straight-forward because the
relative contributions of the stellar and gaseous components to
the total force vary as a function of distance.  For a gas
rich sample of galaxies the derived value of $a_o$ scales as ${H_o}^2$
and for a sample of HSB galaxies dominated by the stellar component 
$a_o \propto {H_o}$.  This is related to a more general
property of MOND:  $a_o$ in its different roles scales differently with
$H_o$.  This fact in itself means that MOND cannot live with any 
distance scale-- to be consistent with MOND, $H_o$ must be in the
range of 60 to 80 km/s-Mpc.  
  
Figure 3 shows two examples of MOND fits to rotation curves.  The dotted and
dashed curves are the Newtonian rotation curves of the stellar and gaseous
disks respectively, and the solid curve is the MOND rotation curve with 
the standard value of $a_o$.  Not only does MOND 
predict the general trend for LSB and HSB galaxies, but it also predicts
the observed rotation curves in detail from the observed distribution
of matter.  This procedure has been carried out for about 100 galaxies, 
76 of which are listed in Table 1; the results are given in terms of the
fitted mass of the stellar disk (in most cases, the only free parameter) and
the implied M/Ls.  

Rotation curves for the entire UMa sample of Sanders \& Verheijen (1998) are
shown in Fig.\ 4 where the curves and points have the same meaning
as in Fig.\ 3.  As noted above, this is a complete and unbiased sample
of spiral galaxies all at about the same distance (here taken to be 15.5
Mpc).  The sample includes both
HSB galaxies (e.g. NGC 3992) and LSB galaxies (e.g. UGC 7089) and covers 
a factor of 10 in centripetal acceleration at the outermost observed
point (Table 1 and Fig.\ 1).  The objects denoted by the asterisk
in Fig.\ 4 are
galaxies previously designated by Verheijen (1997) as being kinematically
disturbed (e.g., non-axisymmetric velocity field caused by bars or 
interactions); the derived rotation curves are less secure for these galaxies.
(Note that fits to the UMa rotation curves using a revised
cluster distance of 18.5 Mpc,
from the Cepheid-based re-calibrated Tully-Fisher relation of 
Sakai et al. 2000, would imply that $a_o$ should be reduced to 
$1.0\times 10^{-8}$ cm/s$^2$.)

This is a fair selection of MOND fits to rotation curves in which the
only free parameter is the M/L of the visible disk (no separate bulge
components were fitted in these cases).  In HSB objects in which the
centripetal acceleration remains large out to the last measure point of the
rotation curve, such as NGC 3953, there is a very small difference between the
Newtonian curve and the MOND curve;  i.e., the observed rotation curve
is reasonably well described by Newtonian theory, as expected when accelerations
are high.  In lower acceleration systems the discrepancy is larger 
(eg. UGC 6667).
In gas-rich galaxies, such as UGC 7089, the shape of the rotation curve
essentially follows from the shape of the
Newtonian rotation curve of the gaseous component, as though the
gas surface density were only scaled up by some factor 
on the order of 10-- a property that has been noticed for spiral galaxies 
(Carignan \& Beaulieu 1989, Hoekstra et al. 2001).  
Emprically, MOND gives the rule which determines the precise scaling.

In
general the MOND curves agree well with the observed curves but there are some
cases where the agreement is less than perfect.  Usually these cases have
an identifiable problem with the observed curve or its interpretation
as a tracer of the radial force distribution.  For example, NGC 4389
is strongly barred, and the neutral hydrogen is contiguous with
the visible disk and bar.  Another example is UGC 6818 which is 
probably interacting with a faint companion at its western edge.
  
Figure 5 shows a less homogeneous sample of rotation curves. These are curves
from the literature based upon observations carried out either at the
Westerbork Radio Synthesis Telescope or the Very Large Array from Sanders
(1996) and McGaugh \& de Blok (1998b), and ranked here in order of decreasing
circular velocity.  These are mostly galaxies with a large angular size so
there are many independent points along the rotation curve.  The selection
includes HSB and LSB galaxies such as NGC 2403 and UGC 128-- two objects with
the same asymptotic rotation velocity ($\approx 130$ km/s).  Here the
general trend, predicted by MOND, is evident:  the LSB exhibits 
a large discrepancy
throughout the galaxy in contrast to the HSB where the discrepancy only becomes
apparent in the outer regions.  In several objects, such as NGC 2403,
NGC 6503, and M33, the quality of the MOND fit is such that, given the
density of points, the fitted curve cannot be distinguished beneath
the observations.  

The most striking
aspect of these studies is the fact that, not only general trends, but also
the details of individual curves are well-reproduced by Milgrom's
simple formula applied to the observed distribution of matter.
In only about 10\% of the roughly 100 galaxies considered in the context
of MOND does the predicted rotation 
curve differ significantly from the observed curve.   

We have emphasized that the only free parameter in these fits is the 
M/L of the visible disk, so one may well ask
if the inferred values are reasonable.  It is useful to consider
again the UMa sample, because
all galaxies are at the same distance and there is K'-band (near infrared)
surface photometry of the entire sample.  Fig.\ 6 shows the M/L in the B-band
required by the MOND fits 
plotted against blue minus visual (B-V) color index (top) and the same for the
K'-band (bottom). We see that in the K'-band M/L $\approx 0.8$ with a 30\%
scatter. In other words, if one were to assume a K'-band M/L of about one at
the outset,  most rotation 
curves would be quite reasonably predicted from the observed light and gas
distribution with no free parameters.
In the B-band, on the other hand, the MOND M/L does appear to be a function
of color in the sense that redder objects have larger M/L values.
This is exactly what is expected from population synthesis models
as is shown by the solid lines in both panels (Bell \& de Jong 2001).
This is quite interesting because there is nothing built into MOND
which would require that redder galaxies should have a higher $M/L_b$;  this
simply follows from the rotation curve fits.

\subsection{Falsification of Modified Newtonian Dynamics with rotation curves}

It is sometimes said that MOND is designed to fit rotation
curves so it is not surprising that it does so.  This is not only a 
trivialization of a remarkable phenomenological success, but it is also
grossly incorrect.  MOND was designed to produce asymptotically flat 
rotation curves
with a given mass-velocity relation (or TF law).  It was not designed to
fit the details of all rotation curves with a single adjustable parameter--
the M/L of the stellar disk
(MOND also performs well on galaxies that are gas-dominated and have no 
adjustable parameter).
It was certainly not designed to provide a reasonable 
dependence of fitted M/L on color.  Indeed, none of the rotation curves
listed in Table 1 were available in 1983; designing a theory to
fit data that are not yet taken is called ``prediction.''

However, modified Newtonian dynamics (MOND)
is particularly vulnerable to falsification by rotation curve
data.  Although there are problems, mentioned above, 
in the measurement and interpretation of velocity field 
and photometric data, MOND should
not ``fail'' too often; especially damaging would be a systematic 
failure for a particular class of objects.  In this regard, 
Lake (1989) has claimed that the value of $a_o$ required to fit rotation  
curves varies with the maximum rotation velocity of the
galaxy in the sense that objects with lower rotation velocities
(and therefore lower luminosity galaxies) require a systematically lower
value of $a_o$.  He supported this claim by rotation curve fits to six
dwarf galaxies with low internal accelerations.  If this were true, then
it would be quite problematic for MOND, implying at
very least a modification of Milgrom's simple formula.
Milgrom (1991) responded to this criticism by pointing out inadequacies
in the data used by Lake-- uncertainties
in the adopted distances and/or inclinations.  Much of the rotation
curve data is also of lower quality than the larger galaxies considered 
in the context of MOND.

Swaters \& Sanders (2002) reconsidered this issue on the basis of extensive 21
cm line observations of a sample of 35 dwarf galaxies (Swaters 1999).  When 
$a_o$ is taken as an additional free parameter,
the effect pointed out by Lake is not seen:  there is no systematic
variation of $a_o$ with the maximum rotation velocity of a galaxy.  There is
a large scatter in the fitted $a_o$, but this is due to the fact that
many dwarf galaxies contain large asymmetries or an irregular distribution
of neutral hydrogen.  Moreover,
the galaxies in this sample have large
distance uncertainties--  the distances in many cases being determined
by group membership.  The mean $a_o$ determined from the entire sample
($\approx 10^{-8}$ cm/s$^2$) is consistent with that
implied by the revised Cepheid-based distance scale.

Given the well-known uncertainties in the interpretation of astronomical
data, it is difficult to claim that MOND is falsified on the basis of a 
single rotation curve.  However, it should generally be possible to 
identify the cause for failures (i.e., poor resolution, bars, interactions,
warps, etc.).  An additional uncertainty is the precise distance
to a galaxy, because, as noted above, the estimated internal acceleration in
a galaxy depends upon its assumed distance. For nearby galaxies, 
such as those of the
Begeman et al. sample, the distances are certainly not known to an accuracy of
better than 25\%.  When the MOND rotation curve is less than a perfect
match to the observed curve,
it is often possible to adjust the distance, within reasonable limits
(i.e., the distance appears as a second free parameter).  
In principle, precise independent distance determinations place more 
severe restrictions on this extra degree of freedom, and are therefore 
relevant to rotation curve tests of MOND.

There are now four galaxies from the original sample of Begeman et al. (1991)
with  Cepheid-based distances.  Three of these-- NGC 2403, NGC 3198, and 
NGC 7331-- have been observed as part of the HST Key Project on
the extragalactic distance scale (Sakai et al. 2000).  For NGC 2403 and
NGC 7331, the MOND rotation curve fits precisely the observed curve at
the Cepheid-based distances.  For NGC 3198, 
MOND clearly prefers a distance which is at least
10\% smaller than the Cepheid-based distance of 13.8 Mpc, 
even with the lower value of $a_o$ implied by the revised distance
scale (Bottema et al. 2002).  Given the likely uncertainties in the
Cepheid method, and in the conversion of a 21 cm line velocity field to
a rotation curve, this cannot be interpreted as problematic for MOND.

A more difficult case is presented by NGC 2841,  
a large spiral galaxy with a Hubble distance of about 10 Mpc \cite{BBS}.
The rotation curve of the galaxy cannot be
fit using MOND if the distance is only 10 Mpc;  MOND, as well as the
Tully-Fisher (TF) law, prefers a distance
of 19 Mpc (Begeman et al. 1991, Sanders 1996).  
A Cepheid distance to this galaxy
has been determined \cite{maceal01} at 14.1 $\pm$ 1.5 Mpc.
At a distance of 15.6 Mpc, the MOND rotation curve of the galaxy still
systematically deviates from the observed curve, and the implied
M/L is 8.5; thus, this galaxy remains
the most difficult case for MOND.  It is none-the-less interesting that
the Cepheid-calibrated TF relation (Sakai et al. 2001) implies
a distance of about 23 Mpc for NGC 2841, and supernova
1999by, if a normal type Ia, would imply a distance of 24 Mpc.  

Overall, the ability of MOND, as ad hoc algorithm, to predict 
galaxy rotation curves with only one free parameter, 
the M/L of the visible disk, is striking.   
This implies that galaxy rotation curves
are entirely determined by the distribution of observable matter.  
Regardless of whether or not MOND is correct as a theory, it does
constitute an observed phenomenology that demands
explanation.  Herein lies a real conundrum for the dark matter
picture.  The natural expectations of dark matter theories for
rotation curves do not look like MOND, and hence fail to reproduce
a whole set of essential observational facts.  The best a dark matter
theory can hope to do is contrive to look like MOND, and hence
reproduce a posteriori the many phenomena that MOND successfully predicts.
This gives one genuine pause to consider how science is suppose to
proceed.

\section{PRESSURE-SUPPORTED SYSTEMS}
\subsection{General properties}

Fig.\ 7 is a log-log plot
of the velocity dispersion versus size for pressure-supported, 
nearly isothermal astronomical systems.  At the bottom left the 
star-shaped points are globular clusters (Pryor \& Meylen 1993,
Trager et al. 1993) and the solid
points are giant molecular clouds in the Galaxy (Solomon et al 1987).  The
group of points (crosses) near the center are high surface
brightness elliptical galaxies (J{\o}rgensen et al 1995a,b, J{\o}rgensen 
1999).  At the upper right, the 
squares indicate X-ray emitting clusters of galaxies from the
compilation by White et al. (1997).  
The triangle-shaped points are the dwarf
spheroidal  systems surrounding the Milky Way (Mateo 1998) and the dashes are
compact dwarf ellipticals (Bender et al. 1992).  
The plotted parameters are taken directly from the relevant observational 
papers.  The  measure of size is not homogeneous-- 
for ellipticals and globular
clusters it is the well-known effective radius, for the X-ray clusters it is
an X-ray intensity isophotal radius, and for the molecular clouds it is an
isophotal radius of CO emission.  The velocity dispersion refers to the
central velocity dispersion for ellipticals and globulars; for the clusters it
is the thermal velocity dispersion of the hot gas;  for the molecular  clouds
it is just the typical line width of the CO emission.  

The parallel lines represent fixed internal accelerations. 
The solid line corresponds to  ${\sigma_l}^2/r = 10^{-8}$ cm/s$^2$ and the 
parallel dashed lines to accelerations 5 times larger or smaller
than this particular value.  It is clear from this diagram
that the internal accelerations in most of these systems are within a
factor of a few of $a_o$.  This also implies that the surface densities in
these systems are near the MOND surface density $\Sigma_m$.  

It is easy to over-interpret such a log-log plot containing different
classes of objects covering many orders of magnitude in each coordinate.
We do not wish to claim a velocity-dispersion-size correlation, although
such a relationship has been previously noticed for individual classes of
objects--  in particular, for molecular clouds (Solomon et al 1987) and 
clusters of galaxies (Mohr \& Evrard 1997).  
Probable pressure-supported systems
such as super-clusters of galaxies (Eisenstein et al. 1996) 
and Ly $\alpha$ forest clouds (Schaye 2001)
are clearly not on this relation, but there are
low density solutions for MOND isothermal objects (Milgrom 1984) that have
internal accelerations far below $a_o$.

It has been noted above that, with MOND, if certain very general conditions
are met,
self-gravitating, pressure-supported systems would be expected to have 
internal accelerations comparable to or less than $a_o$.  
The essential condition is  that these objects should be approximately
isothermal.  It is not
at all  evident how Newtonian theory with dark matter 
can account for the fact that these  
different classes of astronomical objects, covering a large range in size and
located in  very different environments, all appear to have comparable
internal accelerations.
In the context of MOND, the location of an
object in this diagram, above or below the ${\sigma_l}^2/r= a_o$ line, 
is an indicator of the internal dynamics and the
extent to which these dynamics deviate from Newtonian.

Now we consider individual classes of objects on this Figure 7.

\subsection{Luminous elliptical galaxies}

Systems above the solid line in Fig.\ 7, e.g., the luminous elliptical
galaxies, are high surface brightness objects and, in the context of MOND,
would not be expected to show a large mass discrepancy within the bright
optical object.  In other words, if interpreted in terms of Newtonian dynamics,
these objects should not exhibit much need for dark matter within an effective
radius;  this is indicated by analysis of the stellar kinematics in several
individual galaxies (e.g., Saglia et al. 1992).  
MOND isotropic, isothermal spheres have a lower mean internal
acceleration within $r_e$ (about one-quarter $a_o$); i.e., these theoretical
objects lie significantly below the solid line in Fig.\ 7.  This was noted by
Sanders (2000) who pointed out that, to be consistent with their observed
distribution in the $r-\sigma_l$ plane, elliptical
galaxies cannot be represented by MOND isothermal spheres;  these
objects must deviate both from being perfectly isothermal (in the sense that
the velocity dispersion decreases outward) and from perfect
isotropy of the velocity distribution (in the sense that stellar orbits become
radial in the outer regions).

The general properties of ellipticals can be matched  by representing these
objects as high-order, anisotropic polytropic spheres; i.e., objects having a
radial velocity dispersion-density relation of the form $$ {\sigma_r}^2 = A
\rho^{1/n}, \eqno(12)$$ where A is a constant depending upon n, the polytropic
index.  In these models the deviation from isotropy toward more radial orbits 
appears beyond an anisotropy radius, $r_a$.
To reproduce the distribution of ellipticals in the $r_e-\sigma_l$ plane,
it must be the case that $12\le n \le 16$ and that $r_a>r_e$ (i.e., the
radial orbit anisotropy does not extend within an effective radius).
However, the strict homology of models is broken and the mass-velocity
relation given above for isotropic, isothermal spheres (eq.\ 8 )
is replaced by a
more  general relation of the form $${\sigma_l}^4 = q(\Sigma/\Sigma_m)Ga_oM.
\eqno(13)$$  That is to say, for these mixed Newtonian-MOND objects,
the mean surface density enters as an additional parameter; actual
elliptical galaxies would comprise a two parameter family and not a one
parameter family as suggested by the simple MOND $M-\sigma_l$ relation
for a homologous class of objects.

This is consistent with the fact that elliptical galaxies do seem to comprise
a two parameter family as indicated by the small scatter about the
``fundamental plane"-- a relation between the luminosity, effective radius,
and  central line-of-sight velocity dispersion of the form
$L \propto \sigma^a {r_e}^b$ where $a\approx 1.5$ and $b\approx 0.8$
(Dressler et al 1987, Djorgovsky \& Davis 1987).  This has generally been
attributed to the traditional virial theorem combined with a systematic
dependence of M/L upon luminosity (e.g., van Albada et al.
1995).  With MOND,
high order polytropes are Newtonian in the inner region and MONDian beyond an
effective radius.  MOND imposes boundary conditions upon the inner
Newtonian solutions that restrict these solutions to dynamical fundamental
plane, i.e., $M\propto {\sigma}^\alpha {r_e}^\gamma$, where the exponents may 
differ from the Newtonian expectations.  The breaking of homology leads
to a considerable dispersion in the $M-\sigma$ relation owing to a factor of
10 dispersion in $q$ in eq.\ 13.  This is shown in Fig.\ 8a which 
is the $M-\sigma$ relation for the
anisotropic polytropes covering the required range in $n$ and $r_a$. 
A least-square fit yields
$$ M/(10^{11} M_\odot) = 2 \times 10^{-8} [\sigma_d (km/s)]^{3.47}\eqno(14)$$ 
where $\sigma_d$ is the velocity dispersion measured within a circular
aperture of fixed linear size as defined by J{\o}rgensen et al (1995a).
The fact that $q$ in eq.\ 13 
is a well-defined function of mean surface
brightness (roughly a power-law), results in a tighter fundamental plane
relation (Fig.\ 8b) of the form $$M/(10^{11} M_\odot) = {3\times
10^{-5}}[\sigma_d (km\,s^{-1})]^{1.76} [r_e (kpc)]^{0.98} \eqno(15)$$ 
(see Fig.\ 8b).
With $M/L \propto M^{0.17}$ the fundamental plane in its observed form is
reproduced.

The existence of a fundamental plane, in
itself, is not a critical test for MOND because Newtonian theory also
predicts such a relation via the virial theorem.  However, for MOND a single
relation of the form of eq.\ 15 applies for range of non-homologous models-- 
this is due to the underlying dynamics and not to the details of galaxy
formation or subsequent dynamical evolution. A curious aspect of the Newtonian
basis for the fundamental plane is the small scatter in the observed relation
given the likely deviations from homology-- particularly considering a
dynamical history that presumably involves multiple mergers.
Moreover,  Newtonian theory offers no explanation for the
existence of a  mass-velocity dispersion relation (even one with large
scatter).   As noted above, in the context of MOND, a near-isothermal object
with a velocity dispersion of a few hundred km/s will always have a galactic
mass.

The compact dwarf ellipticals (the dashes in Fig.\ 7) have internal
accelerations considerably greater than $a_o$ and a mean surface brightness
larger than $\Sigma_m$.  In the context of MOND, this can only be understood
if these objects deviate considerably from isothermality.  If approximated
by polytropic spheres, these objects would have a polytropic index, $n$,
less than about 10 (for $n\le 5$ Newtonian polytropes no longer
have infinite extent and are not necessarily MONDian objects;  thus, there are 
no restrictions upon the internal accelerations or mean surface densities).
This leads to a prediction:  For such high surface brightness objects the
line-of-sight stellar velocity dispersion should fall more dramatically with
projected radius than in those systems with $<\Sigma> \approx \Sigma_m$.
We might also expect the compact dwarfs not to lie on the fundamental plane
as defined by the lower surface brightness ellipticals.

\subsection{Dwarf spheroidal systems}

With MOND, systems that lie below the solid line in Fig.\ 7, i.e., those
systems with low internal accelerations, would be expected to exhibit larger
discrepancies.  This is particularly true of the dwarf spheroidal systems
with internal accelerations ranging down to 0.1 $a_o$.  On the basis
of the low surface brightness of these systems Milgrom (1983b) predicted
that, when velocity dispersion data became available for the stellar 
component, these systems should have a dynamical mass 10 or more times 
larger than that accounted for by the stars.  
This kinematic data is now available,
and, indeed, these systems, when considered in the context of 
Newtonian dynamics, require a significant dark matter content as is 
indicated by M/L values in the range of 10-100 (Mateo 1998).  

For a spherically-symmetric, isolated, low-density object that is deep 
in the MOND regime, a general mass estimator is given by 
$$M= {81\over 4}{{\sigma^4}\over{Ga_o}}\eqno(16)$$   
where $\sigma$ is the line-of-sight velocity dispersion
(Gerhard \& Spergel 1992, Milgrom 1994b). However, in estimating the
dynamical mass of dwarf spheroidals with MOND, one must consider the fact that
these objects are near  the Galaxy, and the external field effect may be
important. A measure of the degree of isolation of such an object would be
given by $$\eta = {{3\sigma^2/2r_c}\over{{V_\infty}^2/R}}\approx {{g_{i}}\over
{g_{e}}}\eqno(17)$$ where 
$r_c$ is the core radius, $V_{\infty}$ is the asymptotic rotation velocity
of the Galaxy ($\approx 200$ km/s) and $R$ is the galactocentric distance
of the dwarf.  For $\eta<1$ the dwarf spheroidal is dominated by the
the Galactic acceleration field and the external field effect must be
taken into account.  In this case, the dynamical mass is simply given
by the Newtonian estimate with the effective constant of gravity
multiplied by $a_o/g_{e}$.  In the opposite limit, the MOND mass estimator
for a system deep in the MOND limit is given by eq.\ 16.

Gerhard \& Spergel (1992) and Gerhard (1994) have argued that MOND
mass-to-light ratio (M/L) values for dwarf spheroidals, based upon these
estimators,  have a very large range, with, for example,
Fornax requiring a global M/L between 0.2 and 0.3 whereas the UMi dwarf
requires M/L $\approx$ 10-13;  that is to say, although some implied M/L
values are unrealistically low, in other cases, MOND still seems to
require dark matter.  Milgrom (1995) has responded to this criticism by
pointing out that the kinematic data on dwarf spheroidals is very  much
in a state of flux, and when more recent values for $\sigma$ are used
along with the realistic error estimates, the MOND M/L values for 
the dwarf spheroidals span a very reasonable range-- on the order of
one to three.
Velocity dispersion data on dwarf spheroidals compiled by Mateo (1998)
yield the MOND M/L values shown in Fig.\ 9.
In addition, MOND seems to fit the radial variation
of velocity dispersion with a plausible amount of anisotropy
(comparable to or less than required by dark matter) in the
cases for which such data are available (Lokas 2001).
It is clear that, for this class of objects, improved data
gives MOND estimates for M/L values that are generally consistent with 
that expected for standard stellar populations. 

\subsection{Globular clusters and molecular clouds}

The globular clusters in Fig.\ 7 generally lie well above the solid line--
i.e., the internal accelerations are in excess of $a_o$.  This  
implies that these systems should show no significant mass discrepancy
within the half-light radius, as seems to be implied by the very reasonable
M/L values based upon Newtonian dynamics.  For a set of 56 globular clusters 
tabulated by Pryor \& Meylan (1993) the mean Newtonian M/L$_V$ is 2.4 $\pm$ 1.
There are several cases of globular clusters with very low internal
accelerations (for example NGC 6366 having $g_{i}/a_o \approx 0.07$), but 
these are generally cases in which the external Galactic field dominates
(i.e., this object is only 4 kpc from the Galactic Center and $g_e>a_o$).
Periodic tidal shocks may also affect the internal dynamics of the systems
and result in larger core radii than if the systems were completely isolated.

The massive molecular clouds in the Galaxy are a unique class of objects
to be considered in this context, in the sense that they are not 
generally included in discussions of the dark matter problem or global
scaling relations.  However, we see from Fig.\ 7 that the 
internal accelerations within these objects are also roughly comparable to 
$a_o$-- a fact that emerged from the empirically discovered size-line 
width relation for molecular clouds in the Galaxy (Solomon et al 1987).
Milgrom (1989b) noticed that this also implies that the surface density of 
molecular clouds is comparable to $\Sigma_m$-- a property 
too striking to be
entirely coincidental.  The suggested explanation is that molecular
clouds with $g_{i} > a_o \approx g_{e}$ expand via classical
internal two-body evaporation until 
$g_i\approx a_o$ at which point they encounter a barrier to further
evaporation;  this can be seen as a consequence of the fact that an isolated
system in MOND is always bound.  If, alternatively, $g_{i} << a_o\approx
g_{e}$ then there is no barrier to tidal
disruption in the Galaxy.  Thereby, $a_o$ emerges as a preferred internal
acceleration for molecular clouds.  Regions in a galaxy where $g_{e}>a_o$
would, as an additional consequence, lack massive molecular 
clouds (as in the inner 3 kpc of the Galaxy apart from the exceptional 
Galactic Center clouds).
The fact that the molecular clouds lie somewhat below the solid line in
Fig.\ 7 would also suggest that, viewed in the context of Newtonian dynamics,
there should be a dark matter problem for molecular clouds-- that is,
the classical dynamically inferred mass should be somewhat larger than the
mass derived by counting molecules.

Solomon et al (1987) noted that combining the size-line width relation with
the Newtonian virial theorem and an empirical mass-CO luminosity relation for
molecular clouds, results in a luminosity-line width relation that
is analogous to the Faber-Jackson relation for ellipticals.  Viewed in
terms of MOND the corresponding mass-velocity dispersion relation is not just 
analogous-- it is the low mass extrapolation
of the same relation that applies to all pressure supported, nearly 
isothermal systems up to and including clusters of galaxies.  
If one applies eq.\ 8 to objects 
with a velocity dispersion of 4 or 5 km/s (typical of giant molecular clouds), 
then one deduces a mass of a few times $10^5$ M$_\odot$.  No explanation of
global scaling relations for extragalactic objects in terms of dark matter,
can accommodate the extension of the relation to such sub-galactic objects.

\subsection{Small groups of galaxies}

We include small galaxy groups in this section on pressure-supported systems
even though this is more properly a small n-body problem.    
Proceeding from individual galaxies,
the next rung on the ladder is binary galaxies, but it is difficult to
extract meaningful dynamical information about these systems, primarily
because of high contamination by false pairs.  The situation improves
for small groups because of more secure identification with an increasing
number of members. Although uncertainties in the mass determination of 
individual groups remains large, either in the context of Newtonian
dynamics or MOND, it is likely that statistical values deduced for 
selected samples of groups may be representative of the dynamics.

This problem has been considered by Milgrom (1998) who looked primarily at a 
catalog of groups by Tucker et al (1998) taken from the 
Las Campanas Redshift Survey.
The median orbital acceleration of galaxies in this sample of groups is on 
the order of a few percent of $a_o$, so these systems are in the deep 
MOND regime.  Milgrom therefore applied the MOND mass estimator relevant
to this limit, eq.\ 16, and found that median M/L values are reduced from
about 100 based on Newtonian dynamics to around 3 with MOND.  Given 
remaining large uncertainties owing to group identification and
unknown geometry, these results are consistent with MOND.

\subsection{Rich clusters of galaxies}

Clusters of galaxies lie below the ${\sigma_l}^2/r = a_o$
line in Fig.\ 7;  thus, these objects would be expected to exhibit
significant discrepancies.
That this is the case has been known for 70 years (Zwicky 1933),
although the subsequent discovery of hot X-ray emitting gas 
goes some way in alleviating the original discrepancy.  For an isothermal
sphere of hot gas at temperature T, the Newtonian dynamical 
mass within radius
$r_o$, calculated from the equation of hydrostatic equilibrium, is 
$$M_n = {{r_o}\over G} {{kT}\over m} 
\Bigl({{d\,ln(\rho)}\over{d\,ln(r)}}\Bigl),\eqno(18)$$
where $m$ is the mean atomic mass and the logarithmic density gradient
is evaluated at $r_o$.  For the X-ray clusters tabulated by 
White et al. (1997), this Newtonian dynamical mass plotted against
the observable mass, primarily in hot gas, is shown in Fig.\ 10 
(Sanders 1999) where
we see that the dynamical mass is typically about a factor of 4 or 5 larger
than the observed mass in hot gas and in the stellar content of the galaxies.
This rather modest discrepancy viewed in terms of dark matter has led to the
so-called baryon catastrophe-- not enough non-baryonic dark matter in the
context of standard cold dark matter (CDM) cosmology (White et al 1993).  

With MOND, the dynamical mass (assuming an isothermal
gas, eq.\ 8) is given by $$M_m = {(Ga_o)}^{-1} {\Bigl({{kT}\over m}\Bigr)^2}
\Bigl({{d\,ln(\rho)}\over{d\,ln(r)}}\Bigl)^2,\eqno(19)$$ and
this is also shown in Fig.\ 10 again plotted against the observable mass.  
The larger scatter is due to the fact that the temperature and the
logarithmic density gradient enter quadratically in the MOND mass
determination.  Here we see that, using the same value of $a_o$ determined
from nearby galaxy rotation curves, the discrepancy is on average reduced to
about a factor of 2.  The fact that MOND predicts more mass than is
seen in clusters has been pointed out previously in the specific example of 
the Coma cluster (The \& White 1988) and for three clusters with measured
temperature gradients for which the problem is most evident in the central 
regions (Aguirre et al. 2001).

The presence of a central discrepancy is also suggested by strong
gravitational lensing in clusters, i.e., the formation of multiple
images of background sources in the central regions of some clusters
(Sanders 1999).
The critical surface density required for strong lensing is
$$\Sigma_c = {1\over{4\pi}}{{cH_o}\over G}F(z_l,z_s)\eqno(20)$$
where F is a dimensionless function of the lens and source redshifts
that depends upon the cosmological model (Blandford \& Narayan 1992);
typically for clusters and background sources at cosmological
distances $F\approx 10$ (assuming that a MOND cosmology is 
reasonably standard).  It is then evident that $\Sigma_c>\Sigma_m$
which means that strong gravitational lensing always occurs in the
Newtonian regime.  MOND cannot help with any discrepancy detected
by strong gravitational lensing.  Because strong gravitational lensing in
clusters typically indicates a projected mass within 200-300 kpc between
$10^{13}$ and $10^{14}$ M$_\odot$, which is not evidently present in
the form of hot gas and luminous stars, it is clear that there is
a missing mass problem in the central regions of clusters that
cannot be repaired by MOND.  This remaining discrepancy 
could be interpreted as a failure,  or one could say that MOND predicts
that the baryonic mass budget of clusters is not yet complete and that there
is more mass to be detected (Sanders 1999). 
It would have certainly been devastating
for MOND had the predicted mass turned out to be typically less 
than the observed mass in hot gas and stars;  this would be a
definitive falsification. 

There is an additional important aspect of clusters of galaxies regarding
global scaling relations.  As pressure-supported, near isothermal objects,
clusters should lie roughly upon the same $M-\sigma$ relation defined
by the elliptical galaxies.  That this is the case was first pointed out
by Sanders (1994) using X-ray observations of about 16 clusters that
apparently lie upon the extension of the Faber-Jackson relation for elliptical
galaxies.  From eq.\ 8, we find that an object having a line-of-sight velocity
dispersion of 1000 km/s would have a dynamical mass of about 
$0.5 \times 10^{14} M_\odot$ which is comparable to the baryonic mass of
a rich cluster of galaxies.  The fact that the Faber-Jackson relation--
albeit with considerable scatter-- extends from molecular clouds to
massive clusters of galaxies finds a natural explanation in terms of MOND.

\subsection{Super-clusters and Ly$\alpha$ forest clouds}

The largest coherent astronomical objects with the lowest internal 
accelerations are superclusters of galaxies as exemplified by the
Perseus-Pisces filament.  If one assumes 
that this object is virialized in a direction perpendicular to the long
axis of the filament, then a linear mass density ($\mu_o$) for the filament
may be calculated following the arguments given by 
Eisenstein et al. (1996);  by approximating the filament
as an infinitely long, axisymmetric, isothermal cylinder, one finds
$$\mu_o = {{2\sigma^2}\over G}.\eqno(21)$$  Applying this relation to 
Perseus-Pisces, these authors estimate a global M/L in the
supercluster of 450 h-- indicating a serious mass discrepancy.

Milgrom (1997b) has generalized the arguments of Eisenstein et al. (1996) 
and finds that a relation similar to eq.\ 21 holds even if one drops the 
assumptions of axial symmetry and isothermality.  He then derived a MOND
estimator for the line density of a filament:
$$ \mu_o = Q{{{<\sigma_\perp^2>}^2}\over{a_oGr_h}}\eqno(22)$$
where $\sigma_\perp$ is the velocity dispersion perpendicular to
the filament axis, $r_h$ is the half mass radius and Q ($\approx$ 2) depends
upon the velocity anisotropy factor.  Applying this 
expression to Perseus-Pisces Milgrom found a M/L value on the order of 10;
once again the MOND M/L seems to require little or no dark matter, even
on this very large scale.
This is significant because the internal acceleration in this object
is on the order of 0.03 $a_o$, which suggests that the MOND formula
applies, at least approximately, down to this very low acceleration.

The diffuse intergalactic clouds resulting in the Ly$_\alpha$ forest 
absorption
lines in the spectra of distant quasars are also apparently objects with
internal acceleration very much lower than $a_o$.  These have been
considered as self-gravitating objects both in the context of dark matter
(Rees 1986) and MOND (Milgrom 1988).  There is now evidence that
the sizes of
individual absorbers may be as large as 100 kpc, as indicated by observations
of gravitationally lensed quasars and quasar pairs (Schaye 2001).
Given that the widths
of the absorption lines are on the order of 10 km/s, then the 
internal accelerations within these systems may be as small as 
$3\times 10^{-4}$ $a_o$.  Schaye (2001) has argued that the characteristic
sizes of the Ly$\alpha$ clouds are most likely to be comparable to the
Jeans length.  In the context of MOND this would be
$$\lambda_J = \Bigl({{\sigma^4}\over{Ga_o\Sigma}}\Bigr)^{1\over 2}
\eqno(23)$$ where $\Sigma$ is the
mean surface density.  Because the fractional ionization is likely to be
very high ($\Sigma$ is dominated by protons), one finds that
this characteristic size, in terms of MOND should be more on the order of
10 kpc, in contradiction to observations of common lines in quasar pairs.
On this basis, Aguirre et al. (2001) have argued that the 
observed large sizes of the absorbers, perpendicular to the line-of-sight, 
are is inconsistent with the predictions of MOND.  

However, these authors noted that the external field effect provides
a possible escape.  The implied internal 
accelerations of the clouds, if they are roughly spherical with sizes of
100 kpc, are likely to be much smaller than the external acceleration field
resulting from large scale structure, which, as we saw above, is on
the order of several percent $a_o$.  In this case the Jean's length is given
by the traditional Newtonian formula with an effective constant of gravity
which may be 20 or 30 times large than G-- and the sizes can be consistent
with the large observed extent.

\section{THE PHYSICAL BASIS OF MOND}
\subsection{The Bekenstein-Milgrom theory}

In spite of its empirical success, MOND remains a largely {\it ad hoc}
modification of Newtonian gravity or dynamics 
without connection to a more familiar theoretical framework.  This is,
at present, the essential weakness of the idea.   The 
original algorithm (eq.\ 2 or 3) cannot be considered as a theory but as a 
successful phenomenological scheme for which an underlying theory is necessary.
If one attempts to apply Milgrom's original prescription (either as a
modification of gravity or inertia) to an N-body system, then immediate
physical problems arise-- such as, non-conservation of linear momentum (Felten
1984).  

Bekenstein \& Milgrom (1984) recognized this and proposed a 
non-relativistic Lagrangian-based theory of MOND as a modification of
Newtonian gravity.  Given a scalar potential $\phi$, the dynamics of the
theory is contained in
field action $$S_f = -\int d^3r\Bigl[\rho\phi + (8\pi G)^{-1}
{a_o}^2 F\Bigl({{{\nabla\phi}^2}\over{{a_o}^2}}\Bigr)\Bigr]\eqno(24)$$
The particle action takes its standard form.
The field equation, derived, as usual, under the assumption of stationary
action is
$$\nabla\cdot\Bigl[\mu\Bigl({{|\nabla\phi|}\over {a_o}}\Bigr)\nabla\phi
\Bigr] = 4\pi G \rho  \eqno(25)$$
where the function $\mu(x) = dF/dx^2$ must have the asymptotic behavior
required in the simple MOND prescription;  i.e., $F(x^2) = (x^2)^{3/2}$ in
the MOND limit ($x\ll 1$) and $F(x^2) = x^2$ in the Newtonian limit.
The equation of motion for a particle assumes its usual Newtonian form.

Because of the symmetry of the Lagrangian density
to spacetime translations (and to
space rotations), the theory respects the laws of conservation of energy and
(angular) momentum.  Moreover, Bekenstein \& Milgrom demonstrated that, 
in the context
of this theory, the motion of a compound object (e.g., a star or star cluster
in the Galaxy) in an external field is independent of its internal structure
(or internal accelerations) and may be described in terms of its center-of-mass
accelerations; i.e., objects like stars with Newtonian internal accelerations
behave like billiard balls in the external field, even in the MOND limit.
However, the external acceleration field does affect the
internal dynamics of such a subsystem in just the way proposed by Milgrom--
the external field effect.

In addition to enjoying the properties of consistency and conservation, this
modified Poisson equation has an interesting symmetry property.
It is well-known that the usual Poisson
equation is conformally invariant in two spatial dimensions. Conformal
transformations comprise a set of angle preserving co-ordinate transformations
that represent, in effect, a position dependent
transformation of units of length. Many of the well-known equations of physics
(e.g.  Maxwell's equations) are invariant under transformations of this form.
Milgrom (1997a) discovered that there is a non-linear generalization of 
the Poisson
equation that is conformally invariant in D spatial dimensions in the
presence of a source $\rho$.  This is of the form $$\nabla\cdot
\Bigl\{[(\nabla\phi)^2]^{D/2-1}\nabla\phi\Bigr\}=\alpha_D G\rho. \eqno(26)$$  
When D=2 eq.\ 26
becomes the usual Poisson equation, but when D=3 the equation takes the form
that is exactly required for MOND phenomenology.  In other words, the
Bekenstein-Milgrom field equation in the MOND limit is conformally invariant
in three spatial dimensions. The full significance of this result is unclear,
but it should be recalled that much of modern physics rests upon just such
symmetry principles.

The Bekenstein-Milgrom theory is a significant step beyond
Milgrom's original prescription.  Even though the theory is non-covariant,
it demonstrates that MOND can be placed upon a solid theoretical basis and that
MOND phenomenology is not necessarily in contradiction with cherished physical
principles.  Although this is its essential significance, the theory also 
permits a more rigorous consideration of specific aspects of MOND
phenomenology relating to N-body systems-- such as the external field
effect. 

It is, in general, difficult to solve this non-linear 
equation except in cases of high symmetry in which the solution reduces to that
given by the simple algorithm.  Brada \& Milgrom (1995) have derived
analytic solutions for Kuzmin disks and of their disk-plus-bulge
generalizations.  The solution can be obtained in the form of a simple
algebraic relation between the Bekenstein-Milgrom solution and the Newtonian 
field of the same mass
distribution, and this relation can be extended to a wider class of 
disk configurations (e.g., exponential disks) where it holds approximately.
>From this work, it is evident that the simple MOND relation (eq.\ 3) gives
a radial force distribution in a thin disk which is generally within 
10\% of that determined by the Bekenstein-Milgrom field equation.

Brada (1997) has developed a numerical method of solution for N-body problems
based upon a multi-grid technique, and Brada \& Milgrom (1999b) used this
method to consider the important problem of stability of disk galaxies. 
They demonstrated that MOND, as anticipated (Milgrom 1989a), has an effect
similar to a dark  halo in stabilizing a rotationally supported disk against
bar-forming modes. However, there is also a significant difference (also
anticipated by Milgrom 1989a).  In a comparison of MOND
and Newtonian truncated exponential disks with identical rotation curves
(the extra force in the Newtonian case being provided by a rigid dark halo),
Brada and Milgrom found that, as the mean surface density of the disk
decreases (the disk sinks deeper into the MOND regime), the growth rate of the
bar-forming m=2 mode decreases similarly in the two cases.  However, in the
limit of very low surface densities, the MOND growth rate saturates while the
Newtonian growth rate continues to decrease as the halo becomes more dominant.
This effect, shown in Fig.\ 11, may provide an important observational
test:  with MOND, low surface brightness (LSB) galaxies remain marginally
unstable to bar- and spiral-forming modes, whereas in the dark matter case,
halo-dominated LSB disks become totally stable.  Observed LSB galaxies
do have bars and m=2 spirals (McGaugh et al. 1995a).  In the
context of dark matter, these signatures of self-gravity are difficult to
understand in galaxies that are totally halo-dominated (Mihos et al. 1997).

The Brada method has also been applied to calculating various consequences of
the external field effect, such as the influence of a satellite in producing a
warp in the plane of a parent galaxy (Brada \& Milgrom 2000a).  The idea
is that MOND, via the external field effect, offers a mechanism other than
the relatively weak effect of tides in inducing and maintaining warps.  As
noted above, the external field effect is a non-linear aspect of MOND, subsumed
by the Bekenstein-Milgrom field equation; unlike Newtonian theory, even a
constant external acceleration field influences the internal dynamics of a
system.  Brada and Milgrom (2000a) demonstrated that a satellite at the
position and with the mass of the  Magellanic clouds can produce a warp in the
plane of the Galaxy with about the right amplitude and form.  

The response of
dwarf satellite galaxies to the acceleration field of a large parent galaxy
has also been considered (Brada \& Milgrom 2000b).  It was found that
the satellites become more vulnerable to tidal disruption because
of the expansion induced by the external field effect as they approach the
parent galaxy (the effective constant of gravity decreases toward its
Newtonian value).  The distribution of satellite orbits is therefore expected
to differ from the case of Newtonian gravity plus dark matter, although it is
difficult to make definitive predictions because of the unknown initial
distribution of orbital parameters.

Although the Bekenstein-Milgrom theory is an important development for all the
reasons outlined above, it must be emphasized that it remains, at best,
only a clue to the underlying theoretical basis of MOND.  The physical basis
of MOND may lie completely in another direction-- as modified Newtonian
inertia rather than gravity.    However, a major advantage of the theory is
that it lends itself immediately to a covariant generalization as a
non-linear scalar-tensor theory of gravity.  

\subsection{Modified Newtonian Dynamics as a modification of General
Relativity}

As noted above, the near coincidence of $a_o$ with $cH_o$ suggests that 
modified Newtonian dynamics (MOND)
may reflect the influence of cosmology upon particle dynamics or $1/r^2$
attraction.  However, in the context of general relativity, there is no such 
influence of this order, with or without the maximum permissible
cosmological term-- a fact 
that may be deduced from the Birkhoff theorem (Nordtvedt \& Will 1972).
Therefore, general relativity, in a cosmological context, cannot be the 
effective theory of MOND; although, the theory underlying MOND must 
effectively reduce to general relativity in the limit of high accelerations 
(see Will 2001 for current experimental constraints on strong field 
deviations from general relativity).

The first suggested candidate theory (Bekenstein \& Milgrom 1984)
is an unconventional scalar-tensor
theory that is a covariant extension of the non-relativistic
Bekenstein-Milgrom theory.  Here the
Lagrangian for the scalar field is given by  $$L_s = {{{a_o}^2}\over c^4}
F\Bigl[{{\phi_{,\alpha}\phi^{,\alpha} c^4}\over{a_o}^2}\Bigr]
\eqno(27)$$ where F(X) is an
arbitrary positive function of its dimensionless argument. This scalar field,
as usual, interacts with matter jointly with  $g_{\mu\nu}$ via a conformal
transformation of the metric, i.e., the form of the interaction Lagrangian is
taken to be $$L_I = L_I[{\xi(\phi^2)}g_{\mu\nu}...] \eqno(28)$$ where $\xi$ is 
a function of the scalar field (this form preserves weak equivalence where
particles follow geodesics of a physical metric $\hat g_{\mu\nu}
= \xi(\phi^2)g_{\mu\nu}$).  The scalar field action ($\int L_s
\sqrt{-g}d^4x$) is combined with the usual Einstein-Hilbert action of general
relativity and the particle action formed from $L_I$ to give the complete
theory. Thus, the covariant form of the Bekenstein-Milgrom field equation
becomes $$(\mu\phi^{,\alpha})_{;\alpha} = {{4\pi G T}\over {c^4}} \eqno(29)$$
where again $\mu = dF/dX$, and $T$ is the contracted energy-momentum
tensor. The complete theory includes the Einstein field
equation with an additional source term appearing owing to the contribution of
the scalar field to the energy-momentum tensor.  Again we require that  F(X)
has the asymptotic behavior $F(X) \rightarrow X^{3\over 2}$ in the limit where
$X\ll 1$ (the MOND limit) and $F(X)\rightarrow \omega X$ in the limit of 
$X\gg 1$.
Thus, in the limit of large field gradients, the theory becomes a standard
scalar-tensor theory of Brans-Dicke form (Brans \& Dicke 1961); it is
necessary that $\omega>>1000$ if the theory is to be consistent with local
solar system and binary pulsar tests of general relativity (Will 2001). 
Because of the non-standard kinetic Lagrangian (eq.\ 27), Bekenstein (1988)
termed this theory the aquadratic Lagrangian or ``AQUAL'' theory.  

Bekenstein \& Milgrom immediately noticed a physical problem with the theory: 
small disturbances in the scalar field propagate at a velocity faster than
the speed of light in directions parallel to the field gradient in the MOND
regime.  This undesirable property appears to be directly related to the
aquadratic form of the Lagrangian, and is inevitably true in any such theory
in which the scalar force decreases less rapidly than $1/r^2$ in the limit of
low field gradient.  Clearly,  the avoidance of causality paradoxes, if only
in principle, should be  a criterion for physical viability.

The acausal propagation anomaly led Bekenstein (1988a,b) to propose an
alternative scalar-tensor theory in which the field is complex $$\chi =
Ae^{i\phi} \eqno(30)$$ and the Lagrangian assumes its usual quadratic form
$$L_s = {1\over 2}{A^2\phi_{,\alpha}\phi^{,\alpha}} + A_{,\alpha} A^{,\alpha} +
V(A^2) \eqno(31)$$ where $V(A^2)$ is a potential associated with the scalar
field. The unique aspect of the theory is that only the phase couples to 
matter (jointly with $g_{\mu\nu}$ as in eq.\ 28); hence, it is designated
``phase coupling gravitation'' (PCG).  The field equation for the
matter coupling field is then found to be $$(A^2\phi^{,\alpha})_{;\alpha} =
{{4\pi \eta G}\over {c^4}} T\eqno(32) $$ where $\eta$ is a dimensionless
parameter describing the strength of the coupling to matter.  Thus the term
$A^2$ plays the role of the MOND function $\mu$;  $A^2$ is also a
function of $(\nabla\phi)^2$ but the relationship is differential instead of
algebraic.  Bekenstein noted that if $V(A^2) = - kA^6$ precisely the
phenomenology of MOND is recovered by the scalar field $\phi$ 
(here $a_o$ is related to the parameters $k$ and $\eta$).  Because
such a potential implies an unstable vacuum, alternative forms 
were considered by Sanders (1988) who demonstrated that phenomenology
similar to MOND is predicted as long as $dV/dA < 0$ over some range of $A$.
Moreover, in a cosmological context (Sanders 1989), PCG, with the properly 
chosen bare potential,
becomes an effective MOND theory where the cosmological $\dot\phi$ plays the
role of $a_o$.  Romatka (1992) considered PCG as one
of a class of two-scalar plus tensor theories in which the scalar fields couple
in one of their kinetic terms, and demonstrated that, in a
certain limit, Bekenstein's sextic potential theory approaches the original 
AQUAL theory.  This suggests that PCG may suffer from a similar
physical anomaly as AQUAL, and, indeed, Bekenstein (1990) 
discovered that PCG apparently permits no stable background solution for the
field equations-- an illness as serious as that of the acausal propagation
that the theory was invented to cure.

A far more practical problem with AQUAL, PCG, or, in fact, all scalar-tensor
theories in which the scalar field enters as a conformal factor multiplying
the Einstein metric (eq.\ 28), is the failure to predict gravitational lensing
at the level observed in rich clusters of galaxies (Bekenstein \& Sanders
1994).  If one wishes to replace dark matter by a modified theory of gravity
of the scalar-tensor type with the standard coupling to matter, then the scalar
field produces no enhanced deflection of light.  The reason for this is
easy to understand:  in scalar-tensor theories, particles follow geodesics
of a physical metric that is conformally related (as in eq.\ 28) to the usual
Einstein metric.  But Maxwell's equations are conformally invariant
which means that photons take no notice of the scalar field (null geodesics
of the physical and Einstein metrics coincide).  In other words,
the  gravitational lensing mass of an astronomical system should be comparable
to that of the detectable mass in stars and gas and thus much less than the
traditional virial mass. This is in sharp contrast to the observations
(Fort \& Mellier 1994).  

A possible cure for this ailment is a non-conformal relation
between the physical and Einstein metrics-- that is, in transforming
the Einstein metric to the physical metric, a special direction is picked out
for additional squeezing or stretching (Bekenstein 1992, 1993).  
To preserve the isotropy of space, this direction is usually
chosen to be time-like in some preferred cosmological frame as in the
classical stratified theories (Ni 1972).  In this way one may
reproduce the general relativistic relation between the weak-field force on 
slow particles and the deflection of light (Sanders 1997).  
However, the Lorentz invariance of
gravitational dynamics is broken and observable preferred frame 
effects-- such as a polarization of the earth-moon orbit
(M\"uller et al. 1996)-- are inevitable at
some level.  It is of interest that an aquadratic Lagrangian for the
scalar field (similar to eq.\ 27) can provide a mechanism for local
suppression of these effects; essentially, the scalar force may be suppressed
far below the Einstein-Newton force in the limit of solar system accelerations.
On this basis, one could speculate that cosmology is described by a 
preferred frame theory (there is clearly a preferred cosmological frame from 
an observational point-of-view).  Then it may be argued that the 
reconciliation of preferred frame cosmology with general relativistic
local dynamics (weak local preferred frame effects) requires 
MOND phenomenology at low accelerations (Sanders 1997).  However,
any actual theory is highly contrived at this point.

In summary, it is fair to say that, at present, 
there is no satisfactory covariant
generalization of MOND as a modification of general relativity.  But this does
not imply that MOND is wrong any more than the absence of a viable
theory of quantum gravity implies that general relativity is wrong.
It is simply a statement that the theory remains incomplete, and that perhaps
the tinkering with general relativity is not the ideal way to proceed.

\subsection{MOND as a modification of Newtonian inertia}

A different approach has been taken by Milgrom (1994a, 1999) who considers
the possibility that MOND may be viewed
as a modification of particle inertia.  In such theories, at a
non-relativistic level, one replaces the standard particle action
($\int v^2/2\,dt$) by a more complicated object, $A_mS[{\bf r}(t),a_o]$ where
$A_m$ depends upon the body and can be identified with the particle mass,
and $S$ is a functional of the particles trajectory, ${\bf r}(t)$, 
characterized by the parameter $a_o$.  This form ensures weak equivalence. 
Milgrom (1994a) proved that if such an action is to be Galilei invariant and
have the correct limiting behavior (Newtonian as $a_o\rightarrow 0$ and
MONDian as $a_o\rightarrow\infty$), then it must be strongly non-local-- 
i.e., the
motion of a particle at a point in space depends upon its entire past
trajectory.  This non-locality has certain advantages in a dynamical theory; 
for example, because a particles motion depends upon an infinite number of
time-derivatives of the particle's position, the theory does not suffer from
the instabilities typical of higher derivative (weakly non-local) theories.  
Moreover, because of the non-locality, the
acceleration of the center-of-mass of a composite body emerges as the relevant
factor in determining its dynamics (Newtonian or MONDIAN) rather than the
acceleration of its individual components. Milgrom further demonstrated that, 
in the context of such theories,
the simple MOND relation (eq.\ 2) is exact for circular orbits in an
axisymmetric potential (although not for general orbits).

Although these results on the nature of generalized particle actions are
of considerable interest, this, as Milgrom stresses, is not yet a theory
of MOND as modified inertia.  The near coincidence of $a_o$ with $cH_o$ 
suggests that MOND is, in some sense, an effective theory-- that is to say,
MOND phenomenology only arises when the theory is considered in a 
cosmological background (the same may also be true if MOND is due to a
modified theory of gravity).  However, the
cosmology does not necessarily directly affect particle motion;  the same
agent-- a cosmological constant-- may affect both cosmology and dynamics. 
Suppose, for example, that inertia results from the interaction of an
accelerating particle with the vacuum. Suppose further that there is a
non-zero cosmological constant (which is consistent with a range of
observations).  Then because a cosmological constant is an attribute of the
vacuum, we might  expect that it has a non-trivial effect upon particle
inertia at accelerations corresponding to $\approx c\sqrt{\Lambda}$.

The phenomenon of Unruh radiation
(Unruh 1975) provides a hint of how this might happen.  An observer uniformly
accelerating through Minkowski space sees a non-trivial manifestation of
vacuum fields as a thermal bath at  temperature $$kT = {{\hbar}\over {2\pi c}}
a \eqno(33)$$ where $a$ is the acceleration (this is exactly analogous to
Hawking radiation where $a$ is identified with the gravitational acceleration
at the event horizon).  In other words, an observer can gain information about
his state of motion by using a quantum detector.  However, the same
observer accelerating through de Sitter space see a modified thermal bath now
characterized by a temperature $$kT_\Lambda = {{\hbar}\over{2\pi c}}\sqrt{a^2
+ {c^2\Lambda\over 3}} \eqno(34)$$  (Narnhofer et al. 1996, Deser
\& Levin 1997). The presence of a cosmological constant changes the
accelerating observer's perception of the vacuum through the introduction
of a new parameter with units of acceleration ($c\sqrt{\Lambda}$) 
and a magnitude comparable
to $a_o$.  If the observer did not know about the cosmological constant this
would also change his perception of the state of motion.

The Unruh radiation itself is too miniscule to be directly implicated as the
field providing inertia-- it may be, in effect, a tracer of the particle's
inertia.  Milgrom (1999) has suggested that inertia may be what drives
a non-inertial body back to some nearby inertial state-- attempting
to reduce the vacuum radiation to its minimum value.  If that were so
then the relevant quantity, with which to identify inertia, would be
$\Delta T = T_\Lambda - T$.  In that case one could write
$${{2\pi c}\over{\hbar}}k\Delta T = a\mu(a/a_o)\eqno(35)$$
where $$\mu(x) = [1+(2x)^{-2}]^{1/2} - (2x)^{-1}\eqno(36)$$
with $a_o = 2c(\Lambda/3)^{1/2}$.  Inertia defined in this way would have
precisely the two limiting behaviors of MOND.

Again, this is not a theory of MOND as modified inertia, but only a suggestive
line of argument.  To proceed further along this line, a theory of inertia
derived from interaction with vacuum fields is necessary-- something analogous
to induced gravity (Sakharov 1968) in which the curvature of space-time
modifies the behavior of vacuum fields producing an associated action
for the metric field.  If this approach is correct, the free
action of a particle must be derived from the interactions with vacuum fields.

\subsection{Gravitational lensing and no-go theorems}

We have noted above that the phenomenon of gravitational lensing
places strong constraints upon scalar-tensor theories of modified
dynamics-- specifically upon the relation between the physical
metric and the gravitational metric (Bekenstein \& Sanders 1994,
Sanders 1997).  Here we wish to discuss gravitational lensing
in a more general sense because it is the only measurable 
relativistic effect that exists on the scale of galaxies and clusters 
and therefore is generally relevant to
proposed modifications of general relativity that may be only effective
on this scale.  Indeed, several authors
have attempted to formulate no-go theorems for modified gravity on the basis 
of the observed gravitational lensing.  

The first of these was by 
Walker (1994) who considered general metrics of the standard Schwarzschild
form $ds^2 = -B(r)dt^2 + A(r)dr^2 + r^2d\Omega^2$ with the condition
that $A=B^{-1}$ and $B-1 = 2\phi(r)$ where $\phi(r)$ is a general weak field
potential of the form $\phi \propto r^n$.  With these assumptions he 
demonstrated that gravity is actually repulsive for photons if 
$0<n<2$-- i.e., gravitational lenses would be divergent.  
He uses this argument to rule out the Mannheim \& Kazanas (1989) spherical
vacuum solution for Weyl conformal gravity in which $\phi(r)$ contains
such a linear term.  But then Walker went on to consider the mean 
convergence $<\kappa>$ and the variance of the shear $\sigma_\gamma$ in
the context of $\phi(r) \propto \log\,r$ which might be relevant to MOND.
Again with the condition that $A=B^{-1}$, this form of the potential
would imply a mean convergence of $<\kappa> \approx 10^{7}$ while the
observations constrain $\kappa <1$; that is to say, with such an 
effective potential the optical properties of the Universe would be
dramatically different than observed.

The second no-go theorem is by Edery (1999) and is even more sweeping.
Basically the claim is that, again assuming a metric in the standard 
Schwarzschild form with $A=B^{-1}$, any potential 
that yields flat
rotation curves (i.e., $\phi$ falls less rapidly than 1/r), is repulsive
for photons even though it may be attractive for non-relativistic
particles.   This was disputed by Bekenstein et al. (2000) 
who pointed out that solar system tests do not constrain the form of A and B
on a galactic scale, except in the context of a specific gravity theory.
An alternative theory
may exhibit $AB=1$ to high accuracy on a solar system scale 
but $AB\neq 1$ on a galactic scale;
indeed, this is a property of the stratified scalar-tensor theory of 
Sanders (1997) which predicts enhanced deflection in extragalactic sources 
but is also consistent with solar system gravity tests at present levels of 
precision.  

Walker's objection would actually seem to be more problematic for MOND;
here the estimate of enormous mean convergence of due to galactic lenses
is independent of the assumption of $AB=1$.  This
problem has also emerged in a different guise in the galaxy-galaxy lensing
results of Hoekstra et al. (2002).  These results imply
that galaxy halos have a maximum extent; if represented by isothermal
spheres the halos do not extend beyond 470 $h^{-1}$ on average.  With
MOND the equivalent halo for an isolated galaxy would be infinite.
Walker also noted that for MOND to be consistent with a low mean
convergence, the modified force law could only extend to several Mpc
at most, beyond which there must be a return to $r^{-2}$ attraction.

The external field effect provides a likely escape from such objections
in the context of pure (unmodified) MOND.  Basically, no galaxy is isolated.
For an $L^*$ galaxy the acceleration at a radius of 470 kpc is about 
0.02$a_o$.  This is at the level of the external accelerations expected from 
large scale structure.  For lower accelerations, one would expect a
return to a $1/r^2$ law with a larger effective constant of 
gravity.  For this reason, such objections can not be considered
as a falsification of MOND.  It might also be that at very low accelerations
the  attraction really does return to $1/r^2$, as speculated by Sanders (1986),
although there is not yet a compelling reason to modify MOND in this
extreme low acceleration regime.

MONDian gravitational lensing in a qualitative sense is also
considered by Mortlock \& Turner (2001) who, after making  
the reasonable assumption
that the relation between the weak field force and deflection in MOND is
the same as in general relativity (specifically including the factor 2 over the
Newtonian deflection), considered the consequences when the force is calculated
from the MOND equation.  The first result of interest is that the thin lens
approximation (i.e., the deflection in an extended source depends only upon 
the surface density distribution) cannot be made with MOND;  this means that
the deflection depends, in general, upon the density distribution along the 
line-of-sight.  They further pointed out that observations of galaxy-galaxy 
lensing (taking the galaxies to be point masses) is consistent
with MOND, at least within the truncation noted by Hoekstra et al.\ (2002). 
Mortlock \& Turner proposed that a test discriminating between MOND and dark
halos would be provided by azimuthal symmetry of the galaxy-galaxy lensing
signal:  MOND would be consistent with such symmetry whereas halos would not. 
They further noted that gravitational microlensing in the context of MOND would
produce a different signature in the light curves of lensed objects
(particularly in the wings) and that this could be observable in cosmological
microlensing (however, this effect may be limited by the external field of the
galaxy containing the microlensing objects).

In general, in extragalactic lenses such as galaxy
clusters, distribution of shear in background sources 
(and hence apparent dark mass distribution) should be calculable from the 
distribution of observable matter; i.e.,
there should be a strong correlation between
the visible and, in terms of general relativity, the dark mass distribution.   
The theme of correlation between the observable (visible) structure of
a lens and the implied shape of the dark matter distribution was taken
up by Sellwood and Kosowsky (2002) who emphasized that the
observed correlation in position angles between the elongated
light distribution and implied mass distribution (Kochanek 2002) argues
strongly in favor of some form of modified gravity.

It is clear from these discussions that gravitational lensing may 
provide generic tests of the MOND hypothesis vs. the dark matter hypothesis,
and that any more basic theory must produce 
lensing at a level comparable to that of general relativity with dark matter.
This already strongly constrains the sort of theory that may underpin
MOND.

\section{COSMOLOGY AND THE FORMATION OF STRUCTURE}

Considerations of cosmology in the context of MOND might
appear to be premature in the absence of a complete theory.  However, there
are some very general statements that can be made about a possible MOND 
universe independently of any specific underlying theory.  First of 
all, the success of the hot Big Bang with respect to predicting
the thermal spectrum and isotropy of the cosmic microwave background 
as well as the observed abundances of the light isotopes (e.g.,
Tytler et al 2000) strongly implies that
a theory of MOND should preserve the standard model-- at least with 
respect to the evolution of the early hot Universe.  This, in fact, 
may be considered as a requirement on an underlying theory.  Secondly, it
would be contrary to the spirit of MOND if there were cosmologically
significant quantities of non-baryonic cold dark matter 
($\Omega_{cdm}<<1$); i.e., dark matter that clusters on the scale of
galaxies.  This is not to say dark matter is non-existent;
the fact that $\Omega_V$ (luminous matter) is substantially less than 
the $\Omega_b$ (baryons) implied by primordial nucleosynthesis
(Fukugita et al. 1998)
means that there are certainly as-yet-undetected baryons.  Moreover, 
particle dark matter also exist in the form of neutrinos.  
It is now clear from the
detection of neutrino oscillations (Fukuda et al 1998)
that at least some flavors of  neutrinos have mass:  the constraints 
are $.004h^{-2}<\Omega_\nu<0.1h^{-2}$ (Turner 1999)
with the upper limit imposed by the
experimental limit on the electron neutrino mass (3 ev).  However, it would be
entirely inconsistent with MOND if dark matter, baryonic or non-baryonic, 
contributed substantially
to the mass budget of galaxies.  Neutrinos near the upper limit of 3 ev
can not accumulate in galaxies due to the well-known phase space constraints
(Tremaine \& Gunn 1979), but they could collect within and contribute to the 
mass budget
of rich clusters of galaxies (which would not be inconsistent with MOND as
noted above).  But apart from this possibility it is reasonable to assume that 
MOND is most consistent with a purely baryonic universe.

This possibility has been considered by McGaugh (1998) who first
pointed out that, in the absence of cold dark matter (CDM), oscillations should
exist in the present power spectrum of large scale density fluctuations-- at
least in the linear regime.  These oscillations are the relic of the sound
waves frozen into the plasma at the epoch of recombination and are suppressed
in models in which CDM makes a dominant contribution to the mass density
of the Universe (Eisenstein \& Hu 1998).  McGaugh (1999b) further 
considered whether or not a cosmology with $\Omega_m\approx \Omega_b
\approx 0.02h^{-2}$ would be consistent with observations of anisotropies
in the cosmic microwave background, particularly the pattern of acoustic
oscillations predicted in the angular power spectrum (e.g. Hu et al.\ 1997).
McGaugh, using the standard CMBFAST program (Seljak \& Zaldarriaga 1996),
pointed out that, in a purely baryonic Universe with vacuum energy density
being the dominant constituent, the second acoustic peak would be much reduced
with respect to the a priori expectations of the concordance $\Lambda$CDM model
(Ostriker \& Steinhardt 1995).  The reason for this low amplitude is Silk
damping (Silk 1968) in a low $\Omega_m$, pure baryonic universe-- the shorter
wavelength fluctuations are exponentially suppressed by photon diffusion.
When the Boomerang and Maxima results first 
appeared (Hanany et al 2000, Lange et al. 2001), much of the initial
excitement was generated by the unexpected low amplitude of the second
peak.  With $\Omega_{total} = 1.01$ and $\Omega_m=\Omega_b$ (no CDM or
non-baryonic matter of any sort), McGaugh (2000) produced a good match to 
these initial Boomerang results.  
A further prediction is that the third acoustic peak should be even
further suppressed.  There are indications from the more complete analysis 
of Boomerang and Maxima data (Netterfield et al 2001, Lee et al. 2001)
that this may not be the case, but the systematic uncertainties remain large.

The SNIa results on the accelerated expansion of the Universe
(Perlmutter et al 1999) as well as the statistics of gravitational
lensing (Falco et al. 1998) seem to exclude a pure baryonic,
vacuum energy dominated universe, although it is unclear that all of the
systematic effects are well-understood.  It is also possible that a MOND
cosmology differs from a standard Friedmann cosmology in the low-z 
Universe, particularly with regard to the angular size distance-redshift
relation.  At this
point it remains unclear whether these observations require 
CDM. It is evident that such generic cosmological tests 
for CDM relate directly to the viability of MOND.  None-the-less, 
cosmological evidence for dark matter, in the absence of its direct
detection, is still not definitive, particularly considering that a 
MOND universe may be non-Friedmannian.

Can we then, in the absence of a theory, reasonably guess 
what form a MOND cosmology might take?
When a theory is incomplete, the way to proceed is
to make several assumptions-- {\it Ans\"atze}-- in the spirit
of the theory as it stands and determine the consequences.  This has
been done by Felten (1984) and by Sanders (1998) who, following the example
of Newtonian cosmology, considered the dynamics of a finite expanding sphere.
Here it is assumed that the MOND acceleration parameter $a_o$ does not
vary with cosmic time.
The second critical assumption is that, in the absence of a relativistic 
theory, the scale factor of the sphere is also the scale factor of
the Universe, but then an immediate contradiction emerges.
MONDian dynamics of a sphere permits no dimensionless scale factor.
Uniform expansion of a spherical region is not possible, and any such region
will eventually recollapse regardless of its initial density and expansion
velocity (Felten 1984).  In the low acceleration regime, the dynamical
equation for the evolution of the sphere, the MONDian equivalent of
the Friedmann equation, is given by
$$\dot{r} = {u_i}^2 - [2\Omega_m{H_o}^2{r_o}^3a_o]^{1/2}ln(r/r_i)\eqno(37)$$
where $r_i$ is the initial radius of the sphere, $r_o$ is a 
comoving radius, and $u_i$ is the initial 
expansion velocity.  From the form of eq.\ 37 it is obvious that
the sphere will eventually recollapse.  It would also appear 
that a MONDian universe is inconsistent with the cosmological principle.  

However, looking at the Newtonian equations for the dynamics of a
spherical region, one finds that, 
at any given epoch, the acceleration increases linearly with 
radial distance from the center of the sphere.  This means that 
there exists a critical radius given by 
$$r_c = \sqrt{GM_{r_c}/a_o}.\eqno(38)$$
where $M_{r_c}$ is the active gravitational mass within $r_c$. 
Beyond $r_c$ (which is epoch dependent) the acceleration exceeds $a_o$;
therefore, on larger scales, 
the dynamics of any spherical region is Newtonian and the expansion 
may be described by a dimensionless scale factor.  During this Newtonian 
expansion, it would appear possible to 
make the standard assumption of Newtonian cosmology 
that the scale factor of the sphere is identical to the scale
factor of the universe, at least on comoving scales corresponding to
$M_{r_c}$ or larger.  

Making use of the Friedmann equations we find that,
$$r_c = {{2a_o}\over{\Omega_m{H_o}^2}}x^3\eqno(39)$$
during the matter-dominated evolution of the Universe (Sanders 1998).
Here $x$ is the dimensionless scale factor with $x=1$ at present.
Therefore, larger and larger co-moving regions become MONDian as 
the Universe evolves.
Because in the matter-dominated period, the horizon increases as
$r_h\propto x^{1.5}$, it is obvious that at some point in the past, the
scale over which MOND applies was smaller than the horizon scale.
This would suggest that, in the past, whereas small regions may have been
dominated by modified dynamics, the evolution of the Universe 
at large is described by the usual Friedmann models.  In particular,
in the early radiation-dominated universe, $r_c$ is very much smaller than 
the MONDian Jeans length, $\lambda_j \approx (c/H_o)x^{4/3}$ 
(the Hubble deceleration is very large at early times),
so the expectation is that the dynamical history of the early MOND universe
would be identical to the standard Big Bang.

After recombination, the Jeans length of the baryonic component falls very
much below $r_c$ and by a redshift of 3 or 4, $r_c$ approaches the 
horizon scale; the entire Universe becomes MONDian.  Thus we might expect
that the evolution of a post-recombination MOND universe might differ
in significant ways from the standard Friedmann-Lemaitre models particularly
with respect to structure formation.  One could assume that when the
critical radius (eq.\ 39) grows beyond a particular comoving scale,
then the MOND relation (eq.\ 37) applies
for the subsequent evolution of regions on that scale,
and recollapse will occur on a time scale comparable to the  
Hubble time at that epoch.  For a galaxy mass object ($10^{11} M_\odot$,
$r_c\approx 14$ kpc) this happens at a redshift of about 140, and recollapse
would occur on time scale of several hundred million years.  Therefore, we
might expect galaxies to be in place as virialized objects by a redshift
of 10.  It is evident that larger scale structure forms later with the
present turn around radius being at about 30 Mpc.

There are two problems with such a scenario for structure formation.  The
first is conceptual:  In a homogeneous Universe what determines the point 
or points about which such recollapse occurs?  Basically in this picture,
small density fluctuations play no role, whereas we might expect that in the
real world
structure develops from the field of small density fluctuations as in
the standard picture.  The second problem is observational:  this picture
predicts inflow out to scales of tens of megaparsecs;  it would have been quite
difficult for Hubble to have discovered his law if this were true.  

These problems may be overcome in the context of a more physically
consistent, albeit non-relativistic, theory of MOND cosmology 
(Sanders 2001).  Following Bekenstein \& Milgrom (1984), one 
begins with a two-field Lagrangian-based theory of MOND (non-relativistic)
in which one field is to be identified with the usual Newtonian field and
the second field describes an additional MOND force that dominates in
the limit of low accelerations.  The theory embodies two important
properties:  MOND plays no role in the absence of fluctuations (the MOND
field couples only to density inhomogeneities).  This means the basic
Hubble flow is left intact.  Secondly, although the Hubble flow is not
influenced by MOND, it enters as an external field that influences the
internal dynamics of a finite-size region.  Basically, if the Hubble
deceleration (or acceleration) over some scale exceeds $a_o$, the evolution
of fluctuations on that scale is Newtonian.    

Figure 12 shows the growth of fluctuations of different comoving scales 
compared with the usual Newtonian growth.  It is evident
that MOND provides a considerable boost, particularly at the epoch during which
the cosmological constant begins to dominate.  This is due to the fact that the
external acceleration field vanishes at this point, and therefore plays no
role in suppressing the modified dynamics.  This adds a new aspect to an
anthropic argument originally given by Milgrom (1989c):  we are observing the
universe at an epoch during which $\Lambda$ has only recently emerged as the
dominant term in the Friedmann equation because that is when structure
formation proceeds rapidly.  The predicted MOND power spectrum 
(Sanders 2001) is rather similar
in form to the CDM power spectrum in the concordance model, but contains
the baryonic oscillations proposed by McGaugh (1998).  These would be
telling, but are difficult to resolve in large scale structure surveys.

It is the external field effect owing to Hubble deceleration that
tames the very rapid growth of structure in this scenario.  In the
context of the two-field theory, this effect may be turned off,
and then it is only the peculiar accelerations that enter the MOND equation.
This case has been considered by Nusser (2002) who finds extremely rapid
growth to the non-linear regime and notes that the final MOND power 
spectrum is proportional
to $k^{-1}$ independent of its original form.  He found that,
to be consistent with the present amplitude of large scale fluctuations, 
$a_o$ must be reduced by about a factor of 10 over the value determined
from rotation curve fitting.  He has confirmed this with N-body simulations
again applying the MOND equation only in determining the peculiar 
accelerations.  This suggests that the Hubble deceleration should come
into play in a viable theory of MOND structure formation.

All of these conclusions are tentative;  their validity depends
upon the validity of the original assumptions.  
None-the-less, it is evident that MOND is likely to promote
the formation of cosmic structure from very small initial fluctuations;
this, after all, was one of the primary motivations for non-baryonic
cosmic dark matter.

\section{CONCLUSIONS}

It is noteworthy that MOND, as an ad hoc algorithm, can explain many
systematic aspects of the observed properties of bound gravitating 
systems:  (a) the presence of a preferred surface density in 
spiral galaxies and ellipticals-- the Freeman and Fish laws;  
(b) the fact that pressure-supported, nearly isothermal systems ranging from 
molecular clouds to clusters of galaxies are characterized by specific
internal acceleration ($\approx a_o$); (c) the existence of a 
tight rotation velocity-luminosity 
relation for spiral galaxies (the Tully-Fisher law)-- specifically
revealed as a correlation between the total baryonic mass and the 
asymptotically flat rotation velocity of the form $M\propto V^4$;
(d) the existence of a luminosity-velocity dispersion relation in 
elliptical galaxies (Faber-Jackson)-- a relation which extends
to clusters of galaxies as a baryonic mass-temperature relation;  and (e)
the existence of a 
well-defined two parameter family of observed properties, the
fundamental plane, of elliptical galaxies-- objects that have varied
formation and evolutionary histories and non-homologous structure.  Moreover,
this is all accomplished in a theory with a single new parameter with units
of acceleration, $a_o$, that must be within an order of 
magnitude of the cosmologically interesting value of $cH_o$.
Further, many of these systematic aspects of
bound systems do not have any obvious connection to what has been
traditionally called the ``dark matter problem''.  This capacity to
connect seemingly unrelated points is the hallmark of a good theory.    

Impressive as these predictions (or explanations) of systematics may be,
it is the aspect of spiral galaxy rotation curves which is most
remarkable.  The dark matter hypothesis may, in principle, 
explain trends, but the peculiarities of an individual rotation curve 
must result from the unique formation and evolutionary history of that 
particular galaxy.  The fact that there is an algorithm-- MOND--
that allows the form of individual rotation curves to be successfully
predicted from the observed distribution of detectable matter-- stars
and gas-- must surely be seen, at the very least, as a severe challenge for the
dark matter hypothesis.  This challenge would appear to be independent of
whether or not the algorithm has a firm foundation in theoretical physics,
because science is, after all, based upon experiment and observation.
None-the-less, if MOND is, in some sense, correct, then the simple algorithm
carries with it revolutionary implications about the nature of gravity and/or 
inertia-- implications that must be understood in a theoretical sense
before the idea can be unambiguously extended to problems of cosmology and
structure formation.

Does MOND reflect the influence of cosmology on local particle dynamics
at low accelerations?  The coincidence between $a_o$ and $cH_o$ would
suggest a connection.  Does inertia result from interaction of
an acceleration object with the vacuum as some have suggested?  
If so, then one would expect
a cosmological vacuum energy density to influence this interaction.
Are there long-range scalar fields in addition to gravity 
which, in the manner anticipated by the Bekenstein-Milgrom theory, 
become more effective in the limit of low field gradients?  Additional fields
with gravitational strength coupling 
are more-or-less required by string theory, but their influenced must
be suppressed on the scale of the solar system (high accelerations);
otherwise, they would have revealed themselves as deviations from the precise
predictions of general relativity at a fundamental level-- violations of the
equivalence principle or preferred frame effects.  Such suppression can
be achieved via the Bekenstein-Milgrom field equation.

Ideally, a proper theory of MOND would make predictions on a scale 
other than extragalactic;  this would provide the possibility of a more
definitive test.  An example of this is the stratified aquadratic 
scalar-tensor
theory which predicts local preferred frame effects at a level that
should soon be detectable in the lunar laser ranging experiment 
(Sanders 1997).  An additional prediction that is 
generic of viable scalar-tensor theories of MOND 
is the presence of an anomalous acceleration, on the order of $a_o$, 
in the outer solar system.  The reported anomalous acceleration detected
by the Pioneer spacecrafts beyond the orbit of Jupiter (Anderson et al.\ 1998)
is most provocative in this regard, but the magnitude-- $8\times 10^{-8}$
cm/s$^2$-- is somewhat larger than would be naively expected if there is a
connection with MOND. However, this is an example of the kind of test that,
if confirmed, would establish a breakdown of Newtonian gravity or dynamics at
low acceleration.

In a 1990 review on dark matter and alternatives  
(Sanders 1990) wrote 
``Overwhelming support for dark matter would be provided by the laboratory
detection of candidate particles with the required properties, detection
of faint emission from low mass stars well beyond the bright optical image
of galaxies, or the definite observation of `micro-lensing' by
condensed objects in the dark outskirts of galaxies.''
Now, more than
a decade later, a significant baryonic contribution to the halos of
galaxies in the form of ``machos'' or low mass stars seems to have been
ruled out (Alcock et al 2000).  Particle dark matter has been detected 
in the form of neutrinos, but of such low mass-- certainly
less than 3 ev and probably comparable to 0.15 ev (Turner 1999)-- that they 
cannot possibly constitute a significant component of the dark matter-- 
either cosmologically
or on the scale of galaxies.  At the same time, the inferred 
contribution of CDM to the mass budget of the Universe has dropped from 
95\% to perhaps 30\%, and both observational and theoretical problems have 
arisen with the predicted form of halos (Sellwood \& Kosowsky 2001).  
However, all of this has not deterred imaginative theorists from
speculative extrapolations of the standard model to conjure
particles having the properties desired to solve perceived problems with
dark matter halos.  It is surely time to apply Occam's sharp 
razor and seriously consider the
suggestion that Newtonian dynamics may breakdown in the heretofore 
unobserved regime of low accelerations.

\vskip 0.5in
\noindent{ACKNOWLEDGMENTS}

We are very grateful to Moti Milgrom for sharing his deep insight with
us over a number of years, and for many comments and constant encouragement.
We also thank Jacob Bekenstein for his ongoing interest in this
problem and for many hours of stimulating discussions.  
We wish to express our profound gratitude to the observers and
interpreters of the distribution of neutral hydrogen in nearby galaxies-- 
to the radio astronomers of the Kapteyn Institute in Groningen.  In
particular we thank Renzo Sancisi and Tjeerd van Albada
and ``generations''
of Groningen students-- especially, Albert Bosma, Kor Begeman, Adrick Broeils,
Roelof Bottema, Marc Verheijen, Erwin de Blok, and Rob Swaters.

\clearpage

\begin{figure}
\epsfig{file=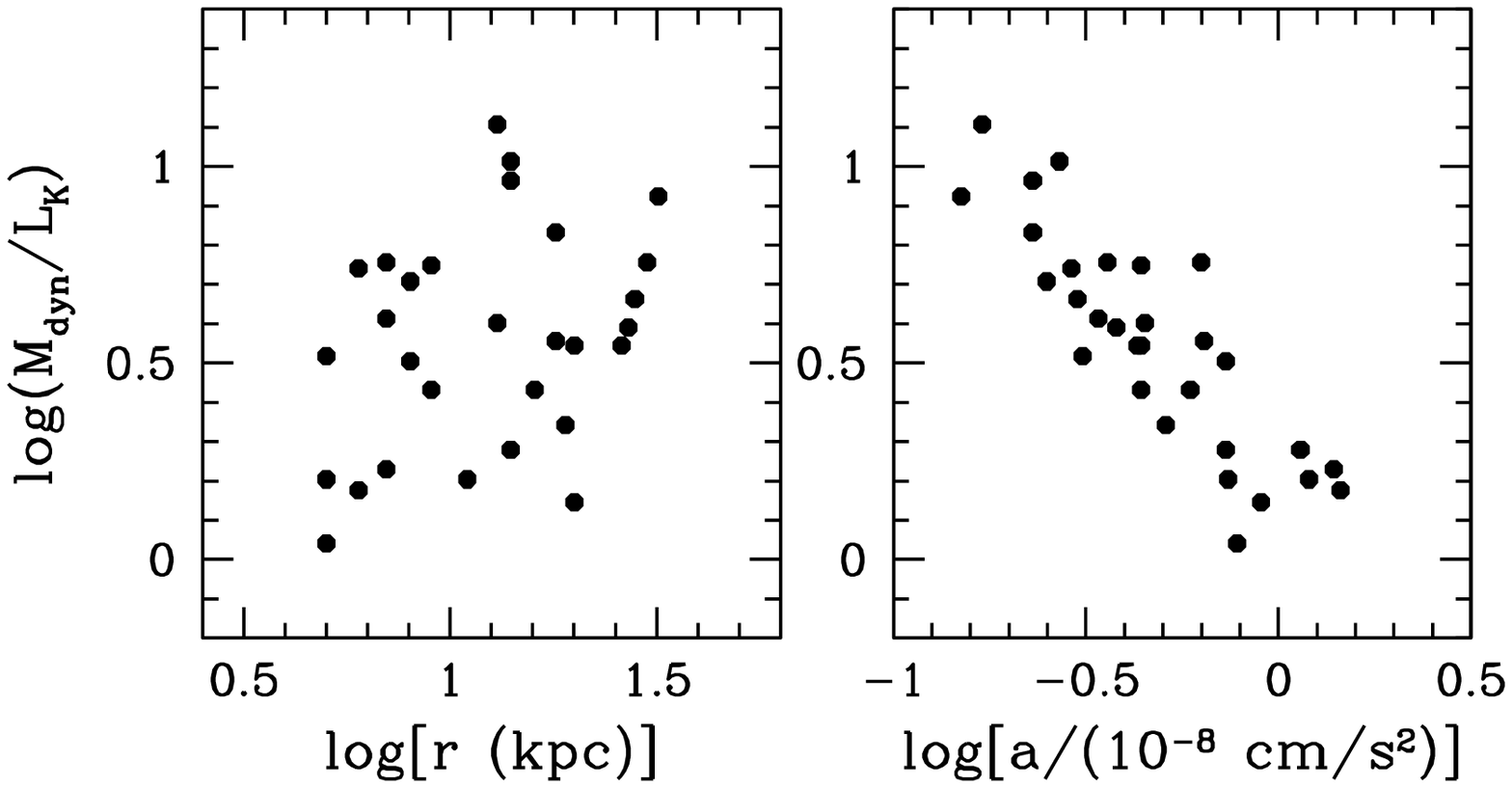,width=14cm}\\
\caption{The global Newtonian mass-to-K'-band-luminosity ratio of Ursa Major
spirals at the last measured point of the rotation curve plotted first against
the radial extent of the rotation curve (left) and then against the centripetal
acceleration at that point (right).}
\end{figure}

\begin{figure}
\epsfig{file=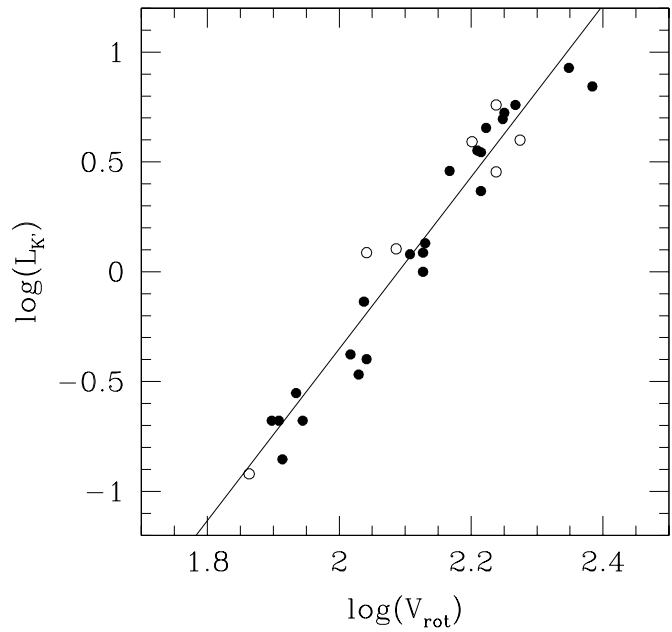,width=10cm}\\
\caption{The near-infrared Tully-Fisher relation of Ursa Major spirals
(\cite{sv98}).  The rotation velocity is the asymptotically constant value.
The velocity is in units of kilometers/second 
and luminosity in $10^{10}$ L$_\odot$.
The unshaded points are galaxies with disturbed kinematics.
The line is a least-square fit to the data and has a slope of $3.9\pm 0.2$}
\end{figure}

\begin{figure}
\epsfig{file=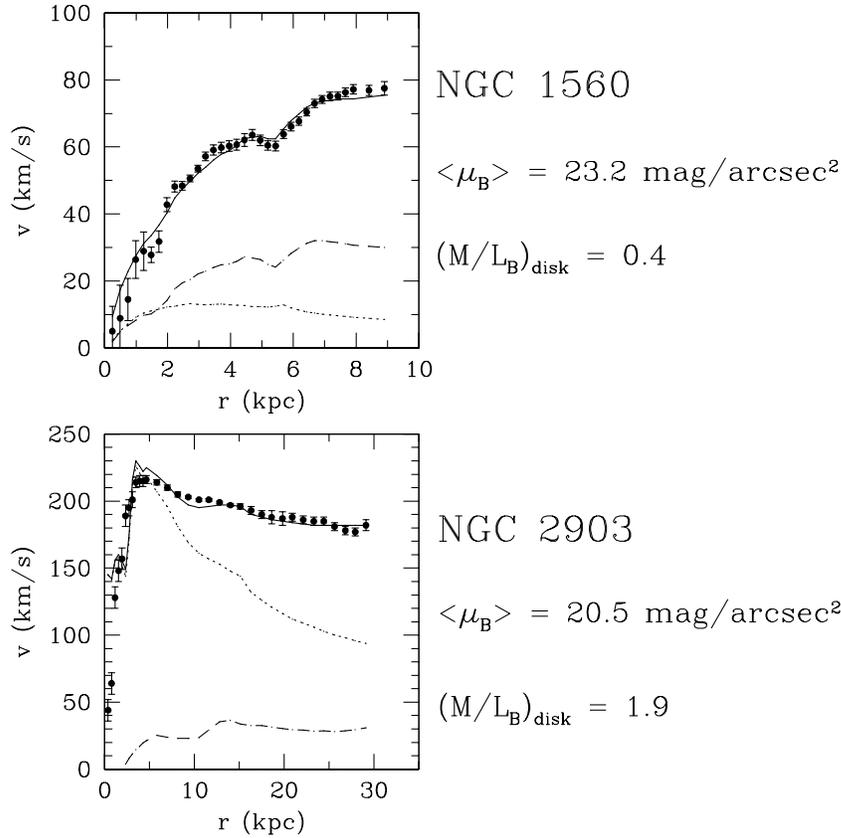,width=12cm}   
\caption{The points show the observed 21 cm line rotation curves of
a low surface brightness galaxy, NGC 1650 (Broeils 1992) and a high
surface brightness galaxy, NGC 2903 (Begeman 1987).  The dotted and dashed
lines are the Newtonian rotation curves of the visible and gaseous
components of the disk and the solid line is the MOND rotation curve  
with $a_o=1.2\times 10^{-8}$ cm/s$^2$-- the value derived from the
rotation curves of 10 nearby galaxies (Begeman et al. 1991). 
The only free parameter is the mass-to-light ratio of the visible
component.}
\end{figure}  

\begin{figure}
\setcounter{figure}{3}
\epsfig{file=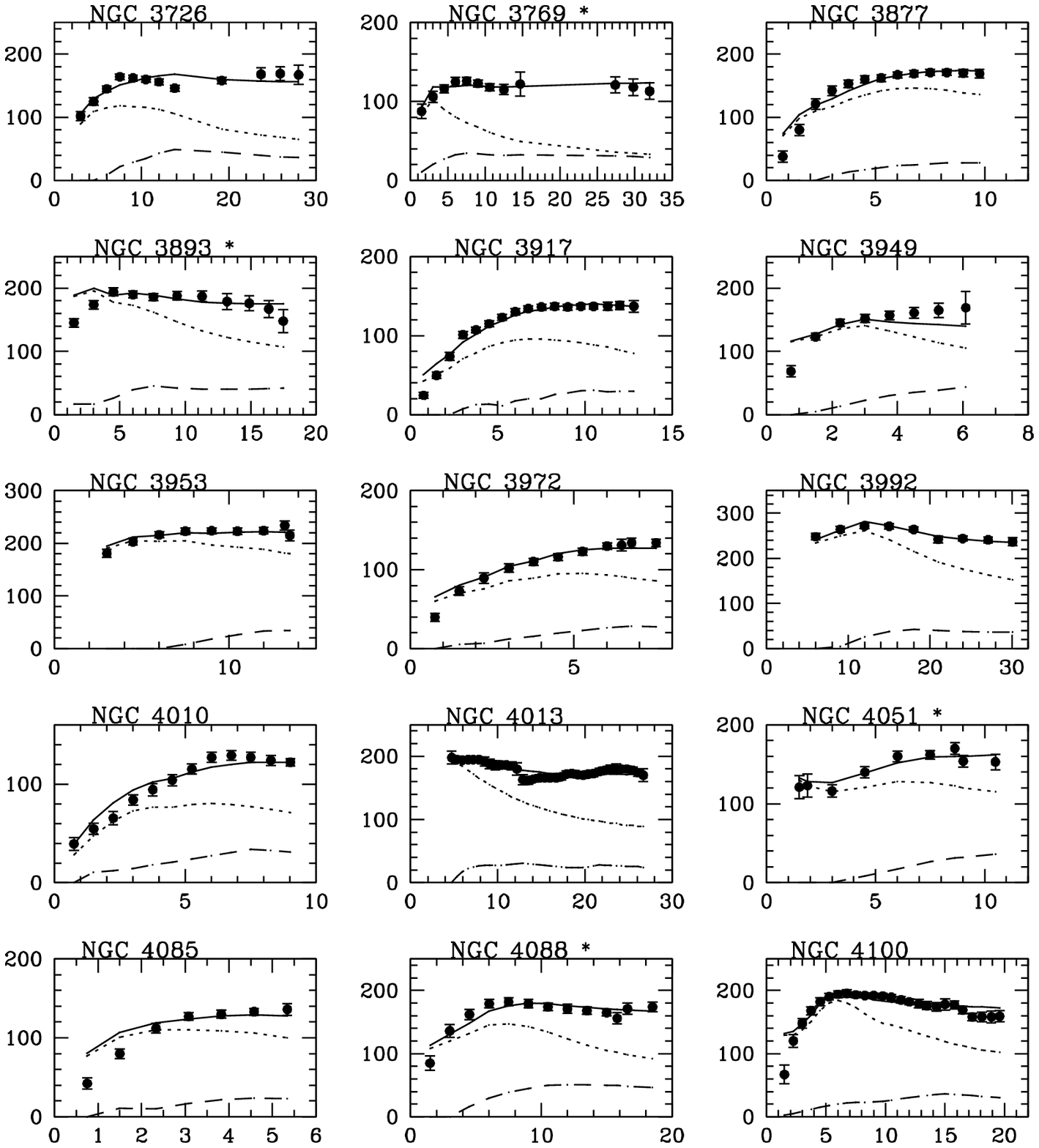,width=14cm}
\caption{}
\end{figure}

\begin{figure}
\setcounter{figure}{3}
\epsfig{file=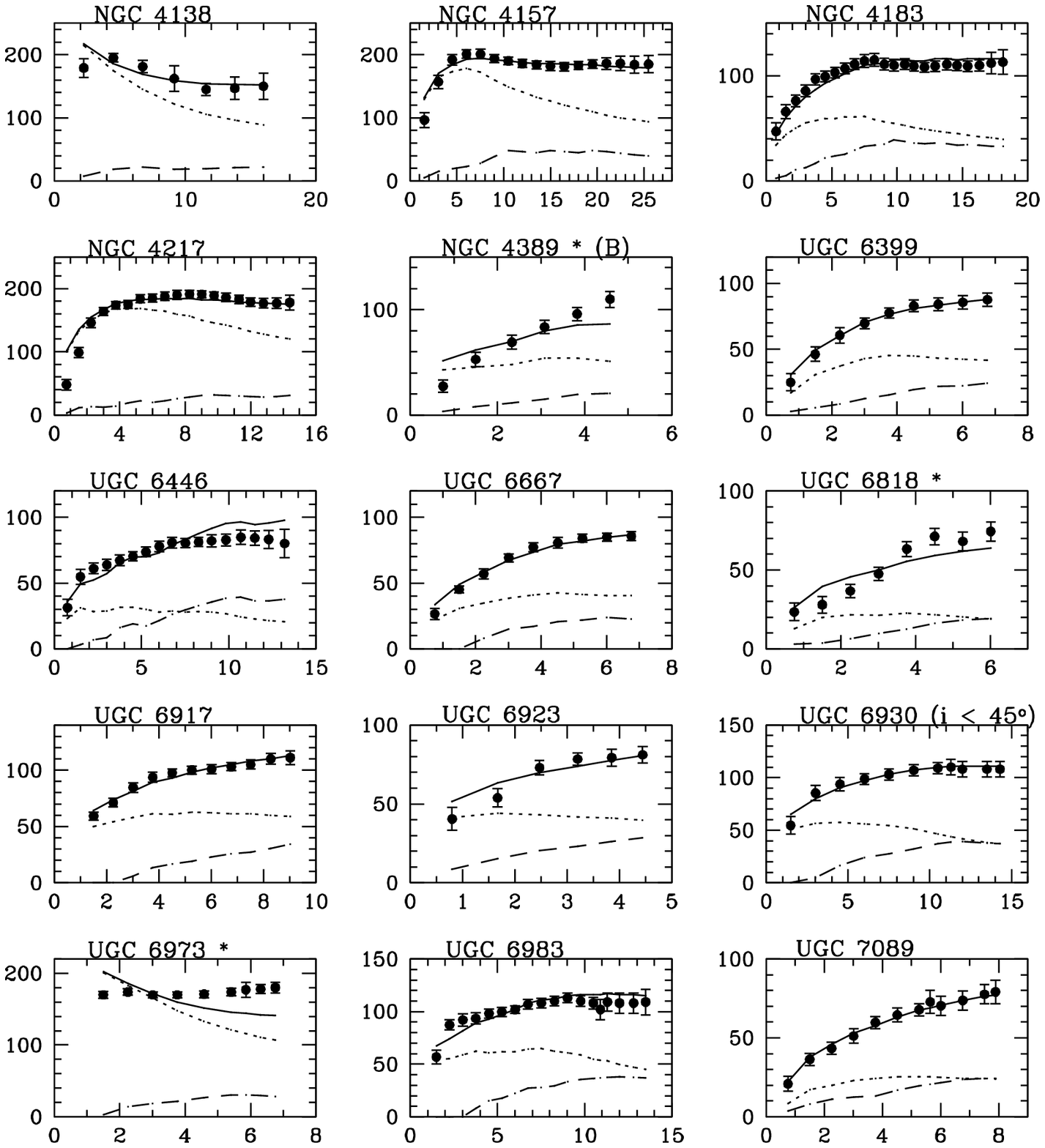,width=14cm}
\caption{ MOND fits to the rotation curves of the Ursa Major galaxies
(Sanders \& Verheijen 1998).
The radius (horizontal axis) is given in kiloparsecs and in all cases the
rotation velocity is in kilometers/second.  The points  and curves 
have the same
meaning as in Fig.\ 3.  The distance to all galaxies is assumed to
be 15.5 Mpc and $a_o$ is the Begeman et al. (1991) value of 
$1.2\times 10^{-8}$ cm/s$^2$. The free parameter of the
fitted curve is the mass of the stellar disk.  If the distance to UMa is 
taken to be 18.6 Mpc, as suggested by the Cepheid-based re-calibration of
the Tully-Fisher relation (Sakai et al. 2000), then $a_o$ must be
reduced to $10^{-8}$ cm/s$^2$.}
\end{figure}

\begin{figure}
\epsfig{file=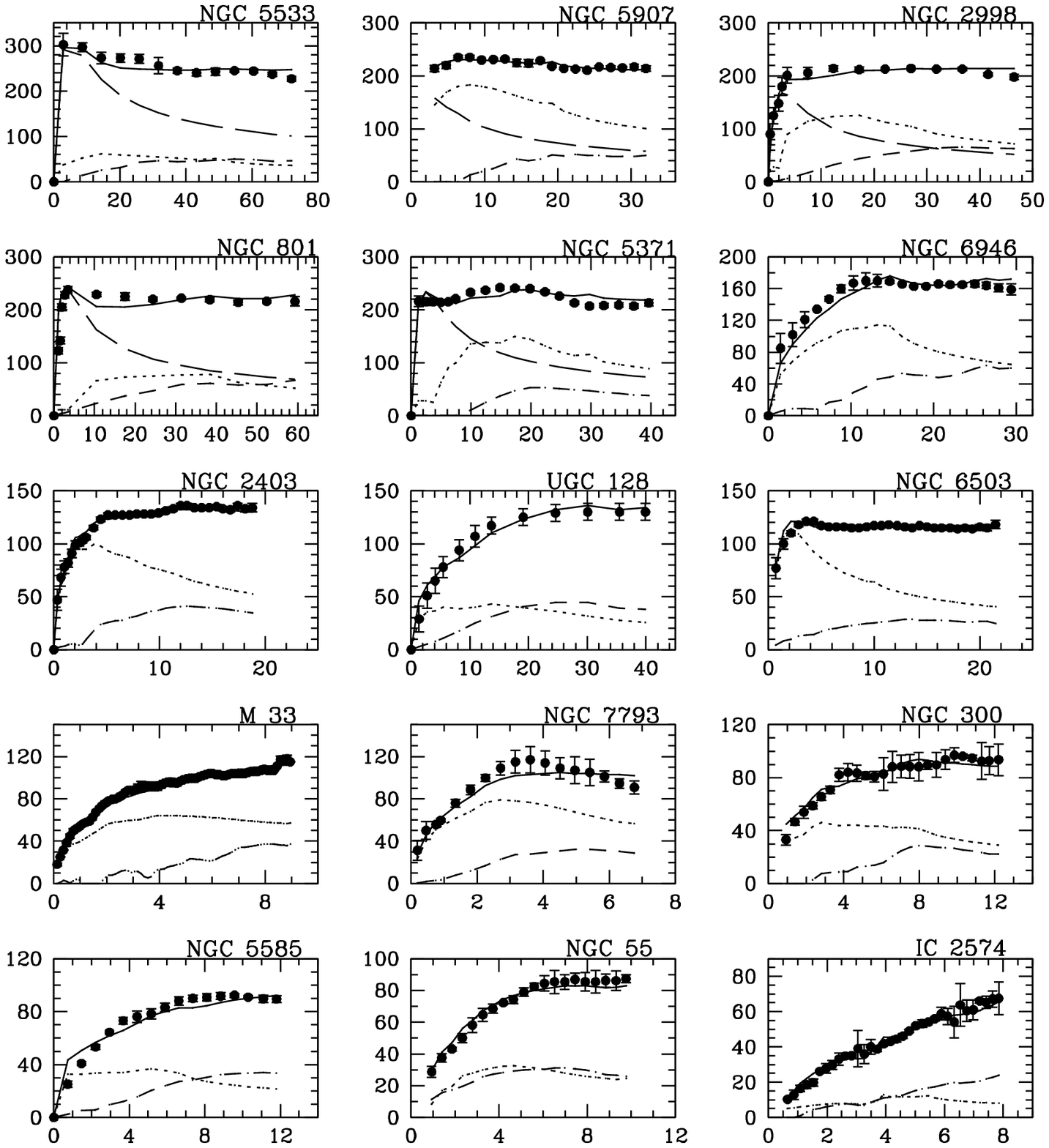,width=14cm}
\caption{MOND fits to the rotation curves of spiral galaxies
with published data, from Sanders (1996) and de Blok \& McGaugh (1998).  The
symbols and curves are as in Fig.\ 4.}
\end{figure}

\begin{figure}
\epsfig{file=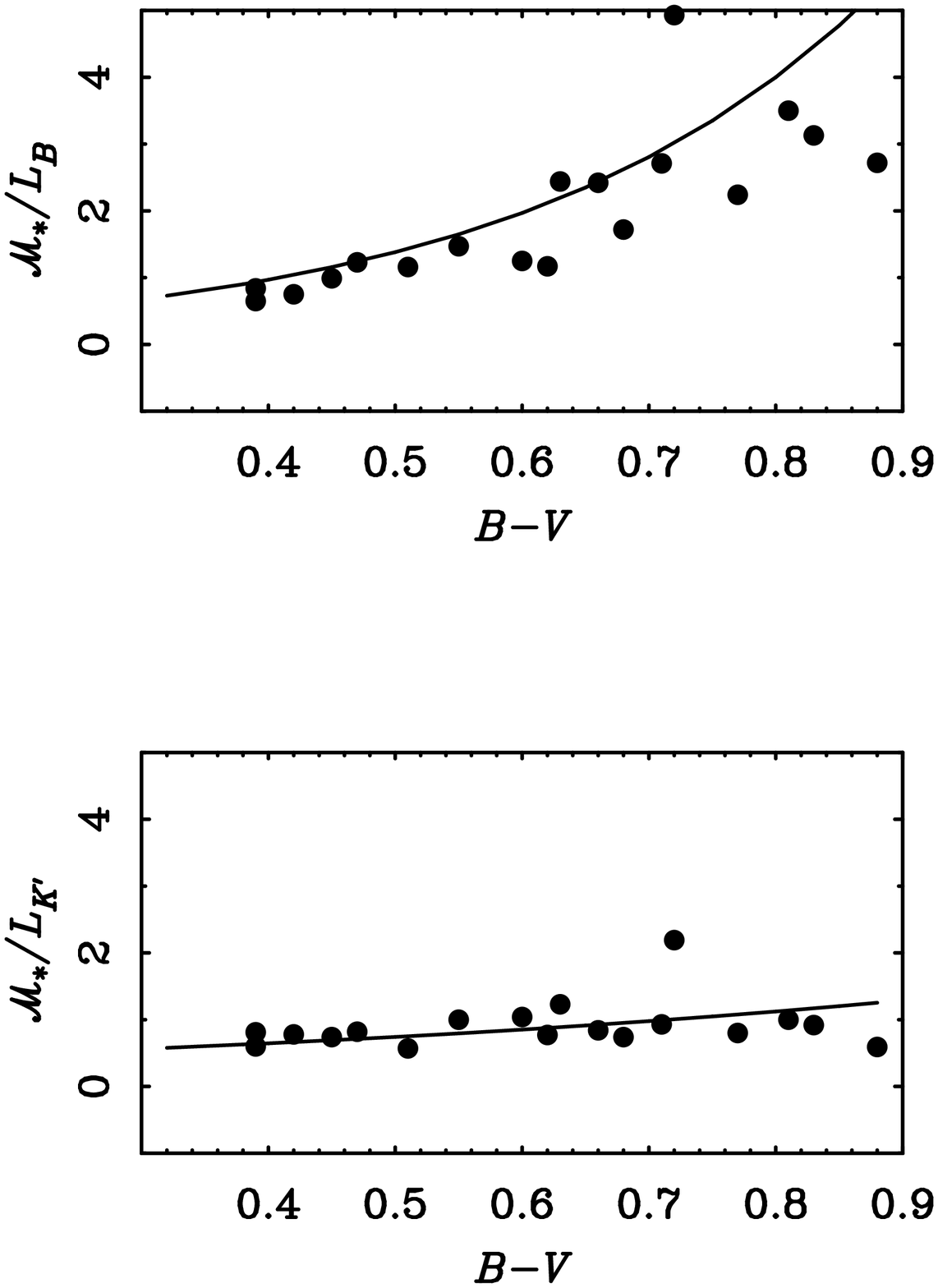,width=12cm}
\caption{Inferred mass-to-light ratios for the UMa spirals (Sanders
\& Verheijen) in the B-band (top) and the K'-band (bottom)
plotted against B-V (blue minus visual) color index.
The solid lines show predictions from
populations synthesis models by Bell and de Jong (2001).}
\end{figure}

\begin{figure}
\epsfig{file=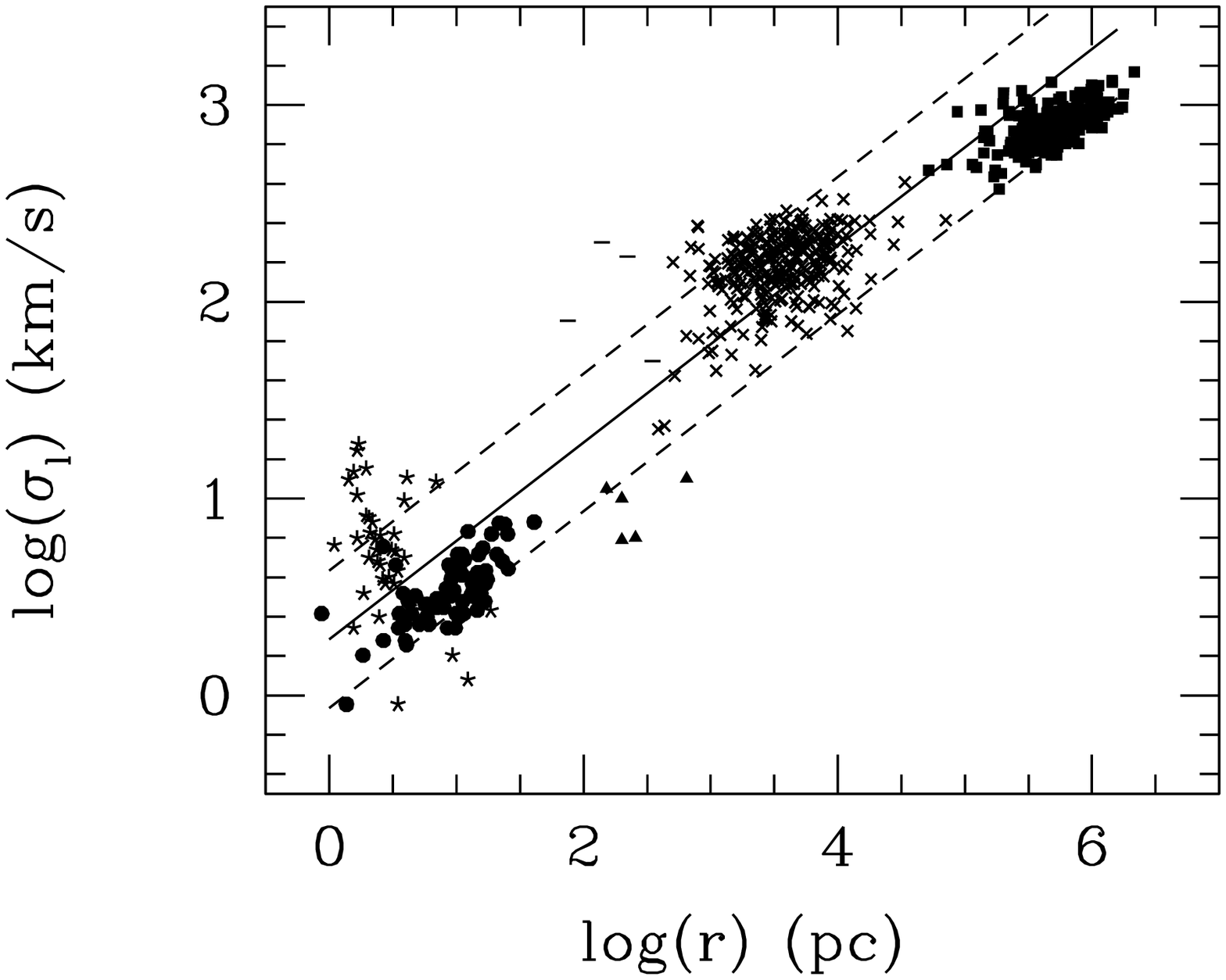,width=14cm}
\caption{The line-of-sight velocity dispersion vs. characteristic
radius for pressure-supported astronomical systems.  The star-shaped
points are globular clusters (Pryor \& Meylen 1993, 
Trager et al. 1993), the points are
massive molecular clouds in the Galaxy (Solomon et al. 1987), 
the triangles are the dwarf spheroidal satellites of the Galaxy
(Mateo 1998), the dashes are compact elliptical galaxies (Bender et al. 
1992), the
crosses are massive elliptical galaxies (J{\o}rgensen et al. 1995a,b, 
J{\o}rgensen 1999),
and the squares are X-ray emitting clusters of galaxies (White, et al.\
1997).  The solid line is shows the relation
$\sigma_l^2/r = a_o$ and the dashed lines a factor of 5 variation about
this relation.}
\end{figure}  

\begin{figure}
\epsfig{file=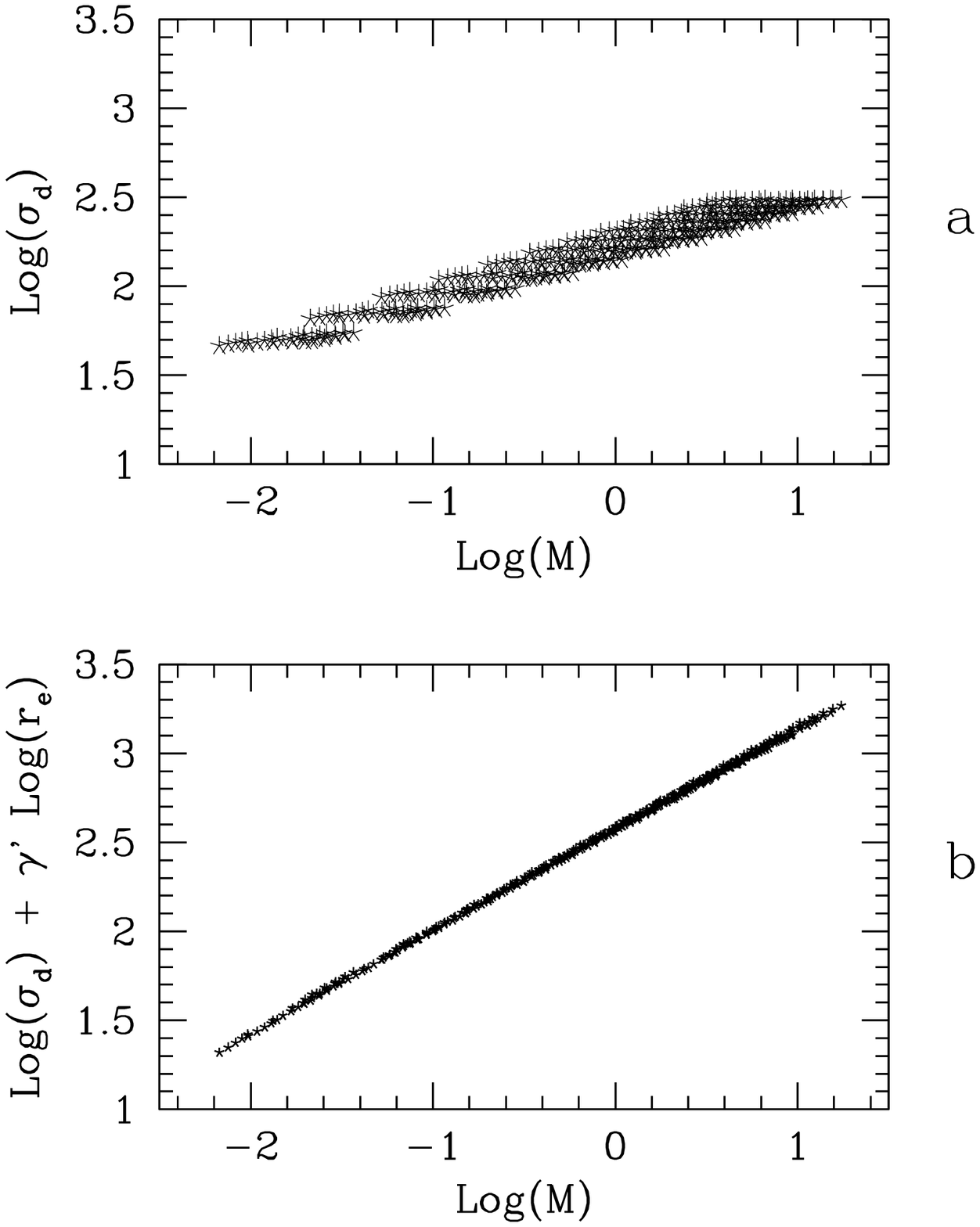,width=12cm}
\caption{a) The mass-velocity dispersion relation for
an ensemble of anisotropic polytropes covering the range necessary
to produce the observed properties of elliptical galaxies.  
Mass is in units of $10^{11}$M$_\odot$ and velocity in kilometers/second. b)
The result of entering a third parameter;  i.e., this is 
the best-fitting fundamental plane relation. log($\sigma_d$ + 
$\gamma'$log($r_e$) is plotted against log($M$) and $\gamma'$ is chosen
to give the lowest scatter ($r_e$ is in kiloparsecs).  
The resulting slope is about 1.76 with $\gamma' = 0.56$.  
>From Sanders (2000).}
\end{figure}

\begin{figure}
\epsfig{file=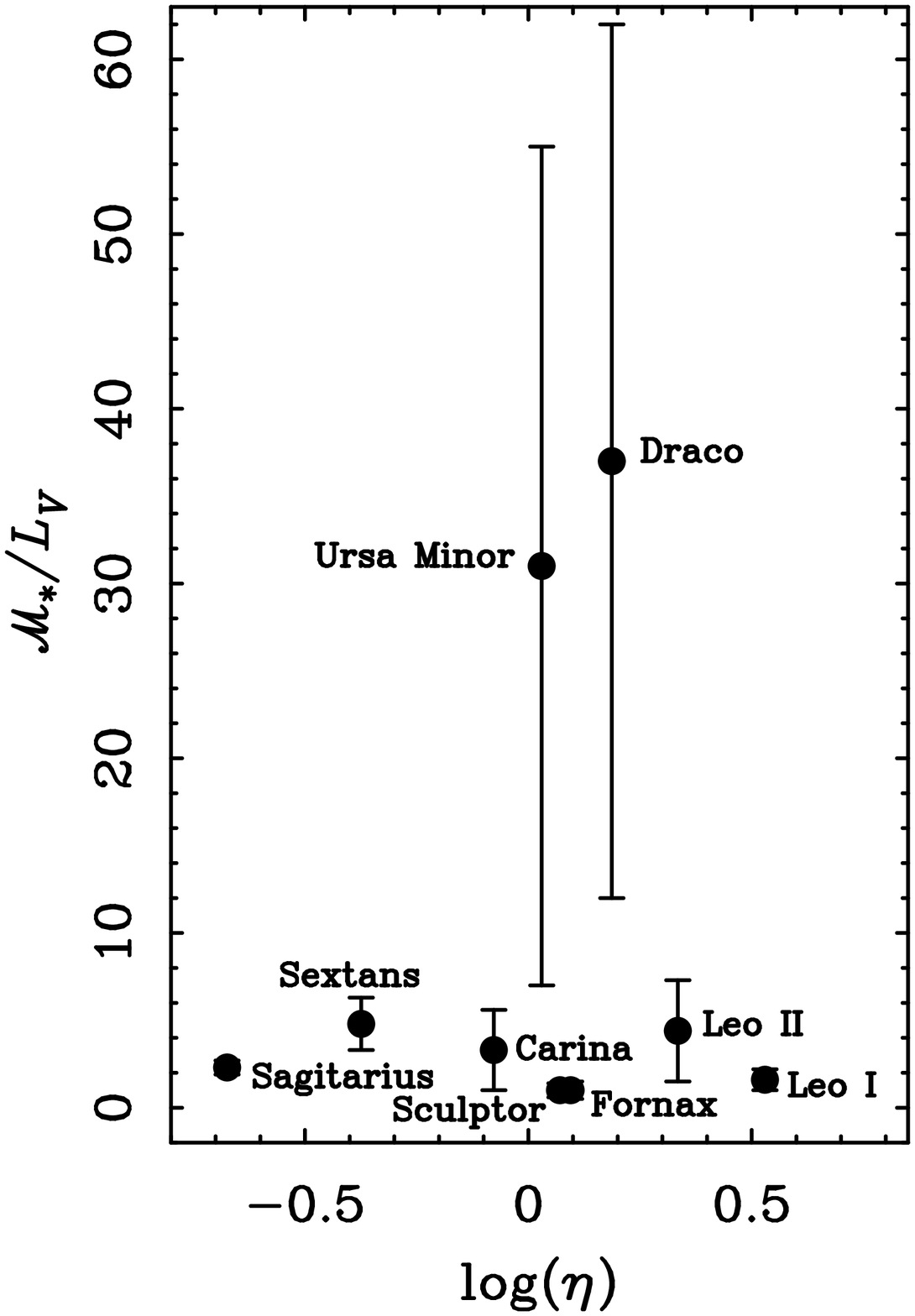,width=12cm}
\caption{The MOND mass-to-light ratio for dwarf spheroidal satellites
of the Galaxy as a function of $\eta$, the ratio of the internal to
external acceleration.  This is the parameter that quantifies
the influence of the Galactic acceleration field (the external field
effect); when $\eta<1$ the object is dominated by the external field.}
\end{figure}

\begin{figure}
\epsfig{file=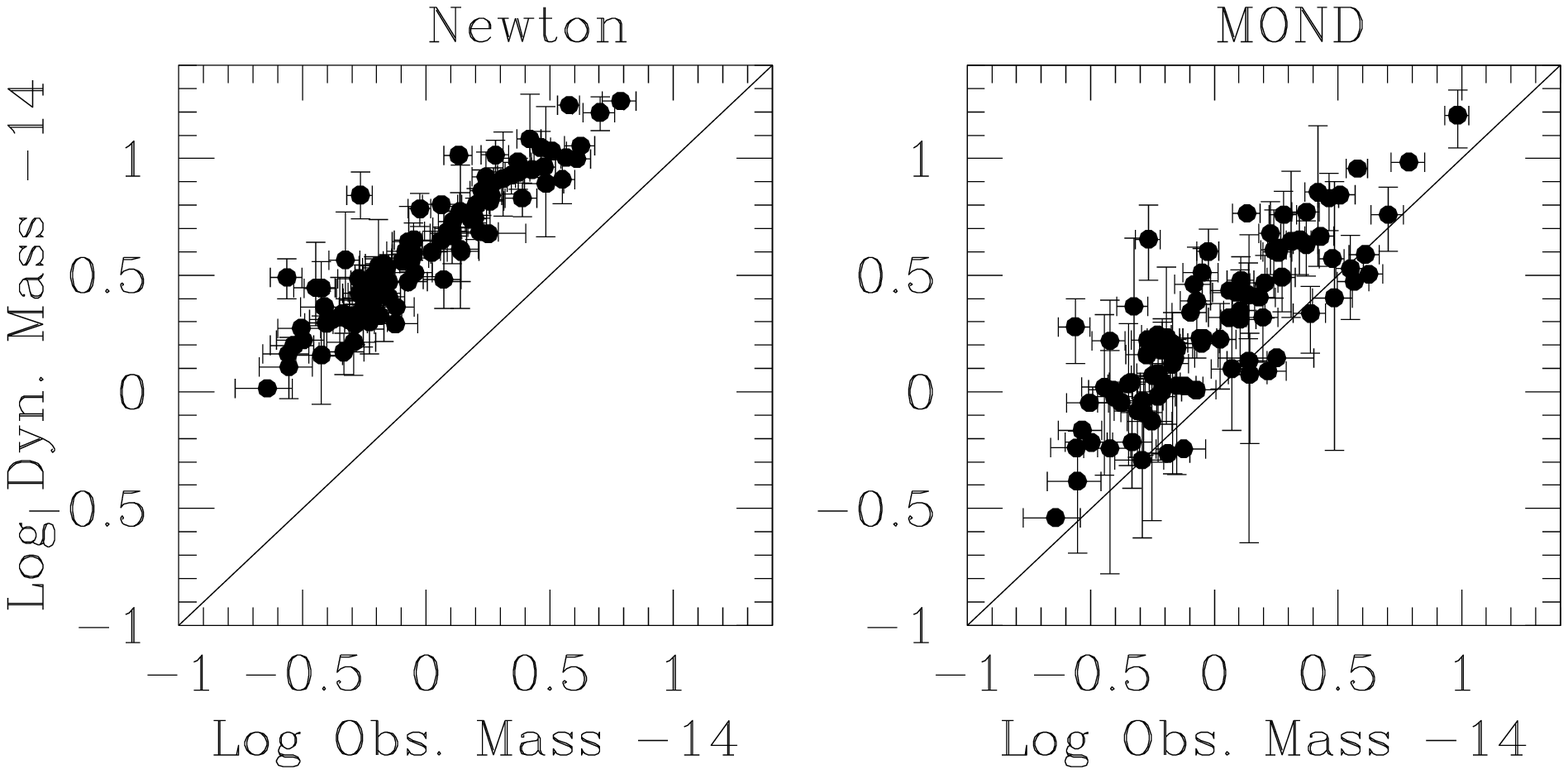,width=14cm}
\caption{({\it Left}) the Newtonian dynamical mass of clusters of 
galaxies within an observed cutoff radius ($r_{out}$) vs. the total 
observable mass in 93
X-ray emitting clusters of galaxies (White et al.\ 1997).
The solid line corresponds to $M_{dyn} = M_{obs}$ (no discrepancy). 
({\it Right}) the MOND dynamical mass 
within $r_{out}$ vs. the total observable mass for the same X-ray emitting 
clusters.  From Sanders (1999).}
\end{figure}

\begin{figure}
\epsfig{file=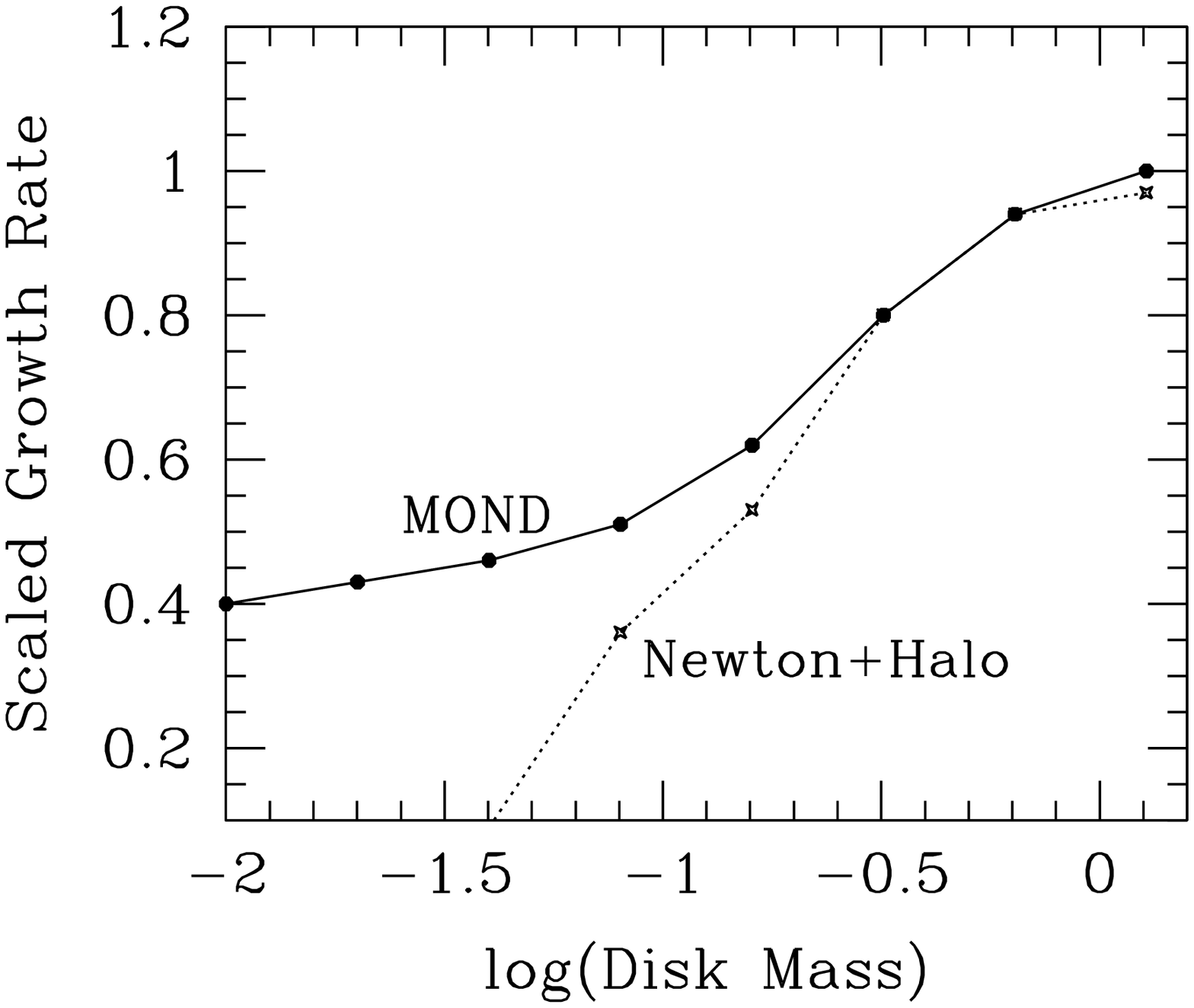,width=10cm}
\caption{The scaled growth rate of the m=2 instability in Newtonian
disks with dark matter (dashed line) and MONDian disks as a function of
disk mass.  In the MOND case, as the disk mass decreases, the surface
density decreases and the disk sinks deeper into the MOND regime.
In the equivalent Newtonian case, the rotation curve is maintained
at the MOND level by supplementing the force with an inert dark
halo. From Brada \& Milgrom 1999b.}
\end{figure}

\begin{figure}
\epsfig{file=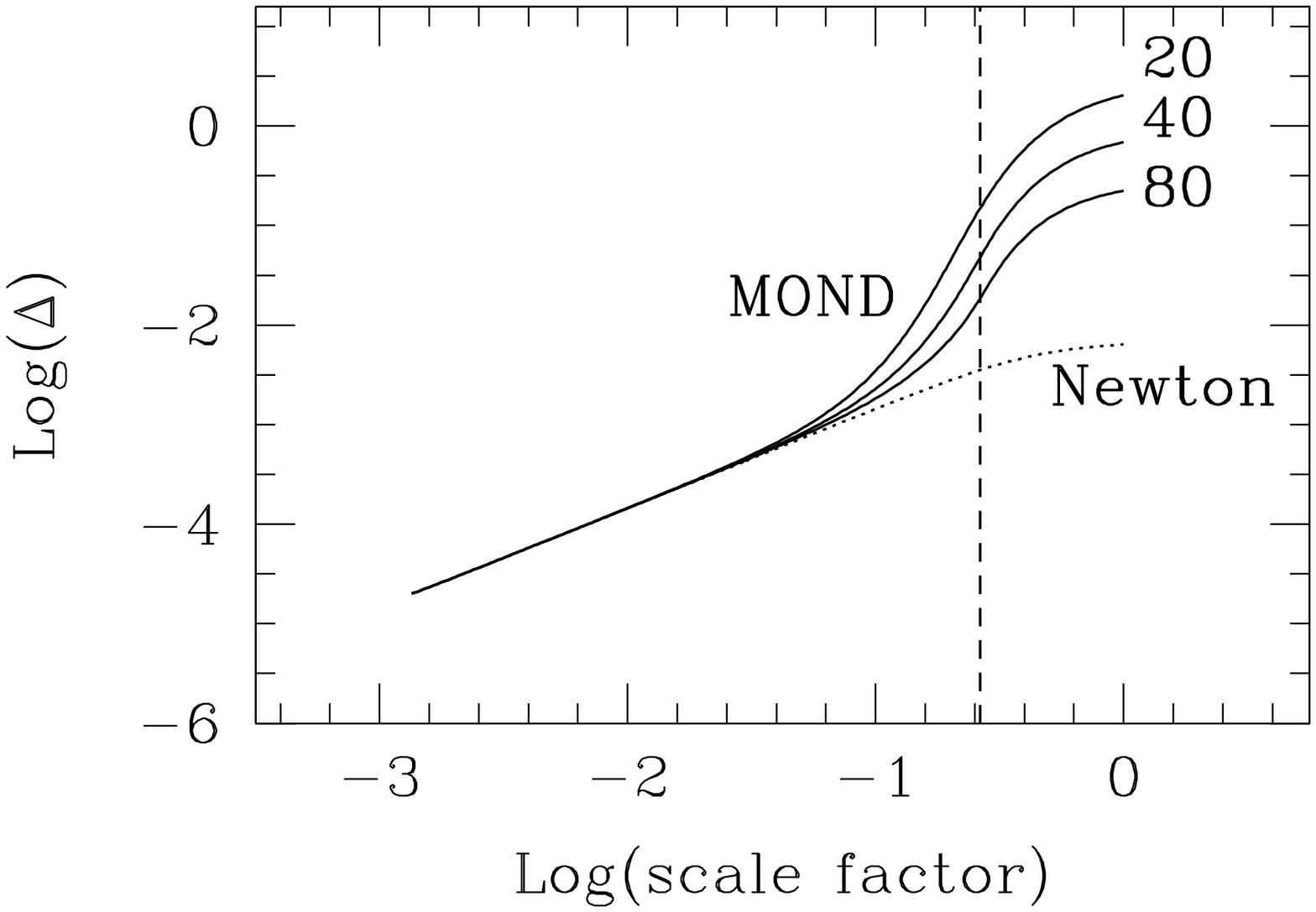,width=12cm}
\caption{The growth of spherically symmetric over-densities in a
low-density baryonic universe as a function of scale factor in the
context of a two-field Lagrangian theory of MOND.  The solid curves
correspond to regions with comoving radii of 20, 40, and 80 Mpc.
The dotted line is the corresponding Newtonian growth.  With MOND  
smaller regions enter the low-acceleration regime sooner and grow
to larger final amplitude.  The vertical dashed line indicates
the epoch at which the cosmological constant begins to dominate
the Hubble expansion.}
\end{figure} 

\setcounter{table}{0}
\begin{table}
\def~{\hphantom{0}}
\noindent {\bf Table 1} Rotation Curve Fits \\
\begin{tabular}{@{}llccccccccl@{}}
\hline
  Galaxy   & Type & $L_B$ & $L_{r}$ & $M_{HI}$ & $V_{\infty}$
 & $M_*$ & $M_*/L_B$ & $M_*/L_{r}$ & ref \\
  (1) & (2) & (3) & (4) & (5) & (6) & (7) & (8) & (9) & (10) \\
\hline
  UGC 2885 & Sbc  & 21.0~~& ---   & 5.0~~& 300 & 30.8~~& 1.5 & --- & 1 \\
  NGC 2841$^a$ & Sb   & ~8.5~~& 17.9~~& 1.7~~& 287 & 32.3~~& 3.8 & 1.8 & 2 \\
  NGC 5533 & Sab  & ~5.6~~& ---   & 3.0~~& 250 & 19.0~~& 3.4 & --- & 1 \\
  NGC 6674 & SBb  & ~6.8~~& ---   & 3.9~~& 242 & 18.0~~& 2.6 & --- & 1 \\
  NGC 3992 & SBbc & ~3.1~~& ~7.0~~& 0.92~& 242 & 15.3~~& 4.9 & 2.2 & 3 \\
  NGC 7331 & Sb   & ~5.4~~& 18.0~~& 1.1~~& 232 & 13.3~~& 2.5 & 0.7 & 2 \\
  NGC 3953 & SBbc & ~2.9~~& ~8.5~~& 0.27~& 223 & ~7.9~~& 2.7 & 0.9 & 3 \\
  NGC 5907 & Sc   & ~2.4~~& ~4.9~~& 1.1~~& 214 & ~9.7~~& 3.9 & 2.0 & 1 \\
  NGC 2998 & SBc  & ~9.0~~& ---   & 3.0~~& 213 & ~8.3~~& 1.2 & --- & 1 \\
  NGC 801  & Sc   & ~7.4~~& ---   & 2.9~~& 208 & 10.0~~& 1.4 & --- & 1 \\
  NGC 5371 & S(B)b& ~7.4~~& ---   & 1.0~~& 208 & 11.5~~& 1.6 & --- & 1 \\
  NGC 5033 & Sc   & ~1.90~& ~3.90~& 0.93~& 195 & ~8.8~~& 4.6 & 2.3 & 1 \\
  NGC 3893$^b$ & Sc   & ~2.14~& ~3.98~& 0.56~& 188 & ~4.20~& 2.0 & 1.1 & 3 \\
  NGC 4157 & Sb   & ~2.00~& ~5.75~& 0.79~& 185 & ~4.83~& 2.4 & 0.8 & 3 \\
  NGC 2903 & Sc   & ~1.53~& ~2.15~& 0.31~& 185 & ~5.5~~& 3.6 & 2.6 & 2 \\
  NGC 4217 & Sb   & ~1.90~& ~5.29~& 0.25~& 178 & ~4.25~& 2.2 & 0.8 & 3 \\
  NGC 4013 & Sb   & ~1.45~& ~4.96~& 0.29~& 177 & ~4.55~& 3.1 & 0.9 & 3 \\
  NGC 3521 & Sbc  & ~2.40~& ---   & 0.63~& 175 & ~6.5~~& 2.7 & --- & 1 \\
  NGC 4088$^b$ & Sbc  & ~2.83~& ~5.75~& 0.79~& 173 & ~3.30~& 1.1 & 0.6 & 3 \\
  UGC 6973$^b$ & Sab  & ~0.62~& ~2.85~& 0.17~& 173 & ~1.69~& 2.7 & 0.6 & 3 \\
  NGC 3877 & Sc   & ~1.94~& ~4.52~& 0.14~& 167 & ~3.35~& 1.7 & 0.7 & 3 \\
  NGC 4100 & Sbc  & ~1.77~& ~3.50~& 0.30~& 164 & ~4.32~& 2.4 & 1.2 & 3 \\
  NGC 3949 & Sbc  & ~1.65~& ~2.33~& 0.33~& 164 & ~1.39~& 0.8 & 0.6 & 3 \\
  NGC 3726 & SBc  & ~2.65~& ~3.56~& 0.62~& 162 & ~2.62~& 1.0 & 0.7 & 3 \\
  NGC 6946 & SABcd& ~5.30~& ---   & 2.7~~& 160 & ~2.7~~& 0.5 & --- & 1 \\
  NGC 4051$^b$ & SBbc & ~2.58~& ~3.91~& 0.26~& 159 & ~3.03~& 1.2 & 0.8 & 3 \\
  NGC 3198$^c$ & Sc   & ~0.90~& ~0.80~& 0.63~& 156 & ~2.3~~& 2.6 & 2.9 & 2 \\
  NGC 2683 & Sb   & ~0.60~& ---   & 0.05~& 155 & ~3.5~~& 5.8 & --- & 1 \\
  UGC 5999$^e$ & Im   & ---   & ~0.13~& 0.25~& 155 & ~0.09~& --- & 0.7 & 4 \\
  NGC 4138 & Sa   & ~0.82~& ~2.88~& 0.14~& 147 & ~2.87~& 3.5 & 1.0 & 3 \\
  NGC 3917 & Scd  & ~1.12~& ~1.35~& 0.18~& 135 & ~1.40~& 1.3 & 1.0 & 3 \\
  NGC 4085 & Sc   & ~0.81~& ~1.22~& 0.13~& 134 & ~1.00~& 1.2 & 0.8 & 3 \\
  NGC 2403 & Sc   & ~0.79~& ~0.98~& 0.47~& 134 & ~1.1~~& 1.4 & 1.1 & 2 \\
  NGC 3972 & Sbc  & ~0.68~& ~1.00~& 0.12~& 134 & ~1.00~& 1.5 & 1.0 & 3 \\
  UGC 128  & Sdm  & ~0.52~& ~0.41~& 0.91~& 131 & ~0.57~& 1.1 & 1.4 & 4 \\
  NGC 4010 & SBd  & ~0.63~& ~1.20~& 0.27~& 128 & ~0.86~& 1.4 & 0.7 & 3 \\
  F568-V1  & Sd   & ~0.22~& ~0.17~& 0.34~& 124 & ~0.66~& 3.0 & 3.8 & 4 \\
  NGC 3769$^b$ & SBb  & ~0.68~& ~1.27~& 0.53~& 122 & ~0.80~& 1.2 & 0.6 & 3 \\
\hline
\end{tabular} \\
\end{table}
\clearpage

\setcounter{table}{0}
\begin{table}
\def~{\hphantom{0}}
\noindent {\bf Table 1} Rotation Curve Fits --- Continued \\
\begin{tabular}{@{}llccccccccl@{}}
\hline
  Galaxy   & Type & $L_B$ & $L_{r}$ & $M_{HI}$ & $V_{\infty}$
 & $M_*$ & $M_*/L_B$ & $M_*/L_{r}$ & ref \\
  (1) & (2) & (3) & (4) & (5) & (6) & (7) & (8) & (9) & (10) \\
\hline
  NGC 6503 & Sc   & ~0.48~& ~0.47~& 0.24~& 121 & ~0.83~& 1.7 & 1.8 & 2 \\
  F568-3   & Sd   & ~0.33~& ~0.27~& 0.39~& 120 & ~0.44~& 1.3 & 1.6 & 4 \\
  F568-1   & Sc   & ~0.28~& ~0.21~& 0.56~& 119 & ~0.83~& 3.0 & 4.0 & 4 \\
  NGC 4183 & Scd  & ~0.90~& ~0.73~& 0.34~& 112 & ~0.59~& 0.7 & 0.8 & 3 \\
  F563-V2  & Irr  & ~0.30~& ---   & 0.32~& 111 & ~0.55~& 1.8 & --- & 4 \\
  F563-1   & Sm   & ~0.14~& ~0.10~& 0.39~& 111 & ~0.40~& 3.0 & 4.0 & 4 \\
  NGC 4389$^b$ & SBbc & ~0.61~& ~1.22~& 0.05~& 110 & ~0.23~& 0.4 & 0.2 & 3 \\
  NGC 1003 & Scd  & ~1.50~& ~0.45~& 0.82~& 110 & ~0.30~& 0.2 & 0.7 & 1 \\
  UGC 6917 & SBd  & ~0.38~& ~0.42~& 0.20~& 110 & ~0.54~& 1.4 & 1.3 & 3 \\
  UGC 6930 & SBd  & ~0.50~& ~0.40~& 0.31~& 110 & ~0.42~& 0.8 & 1.0 & 3 \\
    M 33   & Sc   & ~0.74~& ~0.43~& 0.13~& 107 & ~0.48~& 0.6 & 1.1 & 1 \\
  UGC 6983 & SBcd & ~0.34~& ~0.34~& 0.29~& 107 & ~0.57~& 1.7 & 1.7 & 3 \\
  NGC 247  & SBc  & ~0.35~& ~0.22~& 0.13~& 107 & ~0.40~& 1.1 & 1.8 & 1 \\
  UGC 1230$^e$ & Sm   & ~0.32~& ~0.22~& 0.81~& 102 & ~0.38~& 1.2 & 1.7 & 4 \\
  F574-1$^d$ & Sd   & ---   & ~0.37~& 0.49~& 100 & ~0.26~& --- & 0.7 & 4 \\
  NGC 7793 & Scd  & ~0.34~& ~0.17~& 0.10~& 100 & ~0.41~& 1.2 & 2.4 & 1 \\
  UGC 5005$^e$ & Im   & ---   & ~0.15~& 0.41~& ~99 & ~0.74~& --- & 4.8 & 4 \\
  NGC 300  & Sc   & ~0.30~& ---   & 0.13~& ~90 & ~0.22~& 0.7 & --- & 1 \\
  NGC 5585 & SBcd & ~0.24~& ~0.14~& 0.25~& ~90 & ~0.12~& 0.5 & 0.9 & 1 \\
  NGC 2915$^f$ & BCD  & ~0.04~& ---   & 0.10~& ~90 & ~0.25~& 6.9 & --- & 1 \\
  UGC 6399 & Sm   & ~0.20~& ~0.21~& 0.07~& ~88 & ~0.21~& 1.0 & 1.0 & 3 \\
  NGC 55   & SBm  & ~0.43~& ---   & 0.13~& ~86 & ~0.10~& 0.2 & --- & 1 \\
  UGC 6667 & Scd  & ~0.26~& ~0.28~& 0.08~& ~86 & ~0.25~& 1.0 & 0.9 & 3 \\
  UGC 2259 & SBcd & ~0.10~& ---   & 0.05~& ~86 & ~0.22~& 2.1 & --- & 2 \\
  F583-1   & Sm   & ~0.06~& ~0.06~& 0.24~& ~85 & ~0.11~& 1.7 & 2.0 & 4 \\
  UGC 6446 & Sd   & ~0.25~& ~0.14~& 0.30~& ~82 & ~0.12~& 0.5 & 0.9 & 3 \\
  UGC 6923 & Sdm  & ~0.22~& ~0.21~& 0.08~& ~81 & ~0.16~& 0.8 & 0.8 & 3 \\
  UGC 7089 & Sdm  & ~0.44~& ~0.21~& 0.12~& ~79 & ~0.09~& 0.2 & 0.4 & 3 \\
  UGC 5750$^e$ & SBdm & ---   & ~0.36~& 0.14~& ~75 & ~0.32~& --- & 0.9 & 4 \\
  UGC 6818$^b$ & Sd   & ~0.18~& ~0.12~& 0.10~& ~73 & ~0.04~& 0.2 & 0.3 & 3 \\
  F571-V1$^e$  & Sdm  & ~0.10~& ~0.80~& 0.164& ~73 & ~0.67~& 0.7 & 0.8 & 4 \\
  NGC 1560 & Sd   & ~0.035& ~0.063& 0.098& ~72 & ~0.034& 1.0 & 0.5 & 2 \\
  F583-4   & Sc   & ---   & ~0.071& 0.077& ~67 & ~0.022& --- & 0.3 & 4 \\
   IC 2574 & SBm  & ~0.080& ~0.022& 0.067& ~66 & ~0.010& 0.1 & 0.5 & 1 \\
  DDO 170  & Im   & ~0.016& ---   & 0.061& ~64 & ~0.024& 1.5 & --- & 2 \\
  NGC 3109 & SBm  & ~0.005& ---   & 0.068& ~62 & ~0.005& 0.1 & --- & 2 \\
  DDO 154  & IB   & ~0.005& ---   & 0.045& ~56 & ~0.004& 0.1 & --- & 2 \\
  DDO 168  & Irr  & ~0.022& ---   & 0.032& ~54 & ~0.005& 0.2 & --- & 1 \\
  F565-V2$^e$ & Im   & ~0.023& ~0.019& 0.084& ~51 & ~0.050& 2.2 & 2.7 & 4 \\
\hline
\end{tabular} \\
\end{table}
\clearpage

\noindent EXPLANATION OF COLUMNS OF TABLE 1:

\noindent (1) Galaxy name.  (2) Morphological Type. 
(3) $B$-band luminosity in units of $10^{10}\,L_{\odot}$.
(4) Red band luminosity in units of $10^{10}\,L_{\odot}$.
The precise band used depends on the reference: 
refs.\ 1 \& 2:  $H$-band.  Ref.  3: $K'$-band.  Ref. 4: $R$-band. 
(5) Mass of neutral hydrogen in units of $10^{10}\,M_{\odot}$. 
(6) Asymptotic flat rotation velocity in km s$^{-1}$.
(7) Stellar mass from MOND fit in units of $10^{10}\,M_{\odot}$.
(8) $B$-band stellar mass-to-light ratio in units of $M_{\odot}/L_{\odot}$.
(9) $R$-band stellar mass-to-light ratio in units of $M_{\odot}/L_{\odot}$.
Both $B$ and $R$-band mass-to-light ratios refer only to the stars
(the gas is not included in the mass) and average over disk and bulge
where both components are significant.
See original references for further details. 
(10) References: 1. Sanders (1996).
2. Begeman, Broeils, \& Sanders (1991). 
3. Sanders \& Verheijen (1998).
4. de Blok \& McGaugh (1998). 

\vskip2em

\noindent NOTES FOR TABLE 1:
\vskip0.5em

\noindent\hang $^a$The MOND fit for this galaxy is sensitive to its
distance, preferring $D \approx 19$~Mpc (Sanders 1996) to the Hubble flow
value of $\approx 9$~Mpc.  Macri et al (2001) give a Cepheid distance of
14~Mpc which is marginally tolerable given the uncertainties in this
galaxy's warp (Bottema et al. 2002).

\noindent\hang $^b$Noted as having disturbed kinematics by Verheijen (1997).

\noindent\hang $^c$The MOND fit for this galaxy is sensitive to its distance,
prefering a smaller value than the Cepheid distance of 14.5~Mpc
(Bottema et al.\ 2002).  

\noindent\hang $^d$The original MOND fit for this galaxy
(de Blok \& McGaugh 1998) was not very good.  The 21 cm observations
of this galaxy were severely affected by beam smearing.  

\noindent\hang $^e$Inclination uncertain (de Blok \& McGaugh 1998).

\noindent\hang $^f$Distance uncertain.

\clearpage

\end{document}